\documentclass[usenatbib]{mn2e}

\usepackage{amsmath}
\usepackage{graphicx}
\usepackage{natbib}
\usepackage{color}
\usepackage{xcolor}
\usepackage{times}
\usepackage{multirow}
\definecolor{darkgreen}{rgb}{0.05, 0.4, 0.05}
\definecolor{darkblue}{rgb}{0.1, 0.1, 0.7}

\newcommand{\snia}{SN~Ia}

\def\lesssim{\mathrel{\hbox{\rlap{\hbox{\lower3pt\hbox{$\sim$}}}\hbox{\raise2pt\hbox{$<$}}}}}
\def\gtrsim{\mathrel{\hbox{\rlap{\hbox{\lower3pt\hbox{$\sim$}}}\hbox{\raise2pt\hbox{$>$}}}}}

\newcommand{\tdfdr}{2dFDR}

\newcommand{\targets}{28,504\ } 
\newcommand{\ztotalAll}{17,893\ }
\newcommand{\ztotal}{14,693\ }
\newcommand{\zhost}{2,566\ }

\title[OzDES three-year results]{OzDES multifibre spectroscopy for the Dark Energy Survey:\\ Three year results and first data release}

\author[DES Collaboration]{
\parbox{\textwidth}{
\large
M.~J.~Childress$^{1}$,
C.~Lidman$^{2,3}$,
T.~M.~Davis$^{4,3}$,
B.~E.~Tucker$^{5,3}$,
J.~Asorey$^{4,3}$,
F.~Yuan$^{3,5}$,
T. M. C.~Abbott$^{6}$,
F.~B.~Abdalla$^{7,8}$,
S.~Allam$^{9}$,
J.~Annis$^{9}$
% ...
\begin{center} (DES Collaboration) \end{center}
}
\vspace{0.4cm}
\\
\parbox{\textwidth}{
\scriptsize
$^{1}$ School of Physics and Astronomy, University of Southampton,  Southampton, SO17 1BJ, UK\\
$^{2}$ Australian Astronomical Observatory, 105 Delhi Road, North Ryde, NSW 2113, Australia\\
$^{3}$ ARC Centre of Excellence for All-sky Astrophysics (CAASTRO)\\
$^{4}$ School of Mathematics and Physics, University of Queensland,  Brisbane, QLD 4072, Australia\\
$^{5}$ The Research School of Astronomy and Astrophysics, Australian National University, ACT 2601, Australia\\
$^{6}$ Cerro Tololo Inter-American Observatory, National Optical Astronomy Observatory, Casilla 603, La Serena, Chile\\
$^{7}$ Department of Physics \& Astronomy, University College London, Gower Street, London, WC1E 6BT, UK\\
$^{8}$ Department of Physics and Electronics, Rhodes University, PO Box 94, Grahamstown, 6140, South Africa\\
$^{9}$ Fermi National Accelerator Laboratory, P. O. Box 500, Batavia, IL 60510, USA\\
% ...
}
}

\begin{document}
\maketitle
\begin{abstract}
  We present results for the first three years of OzDES, a six-year
  program to obtain redshifts for objects in the Dark Energy Survey
  (DES) supernova fields using the 2dF fibre positioner and AAOmega
  spectrograph on the Anglo-Australian Telescope.  OzDES is a
  multi-object spectroscopic survey targeting multiple types of
  targets at multiple epochs over a multi-year baseline, and is one of
  the first multi-object spectroscopic surveys to dynamically include
  transients into the target list soon after their discovery. At the
  end of three years, OzDES has spectroscopically confirmed almost 100
  supernovae, and has measured redshifts for 17,000 objects, including
  the redshifts of \zhost supernova hosts. We examine how our ability to
  measure redshifts for targets of various types depends on
  signal-to-noise, magnitude, and exposure time, finding that our
  redshift success rate increases significantly at a signal-to-noise
  of 2 to 3 per 1-{\AA}ngstrom bin. We also find that the change in
  signal-to-noise with exposure time closely matches the Poisson limit
  for stacked exposures as long as 10 hours. We use these results to
  predict the redshift yield of the full OzDES survey, as well as the
  potential yields of future surveys on other facilities such as
  4MOST, PFS, and MSE.  This work marks the first OzDES data release,
  comprising \ztotal redshifts.  OzDES is on target to obtain over
  30,000 redshifts over the six-year duration of the survey, including
  a yield of approximately 5,700 supernova host-galaxy redshifts.

\end{abstract}


\section{Introduction}
\label{sec:intro}

Wide-field, multi-object spectroscopy (MOS) has revolutionized our
understanding of cosmology, galaxy evolution and Galactic structure.
Historically, most large MOS campaigns have used target lists that are
defined well in advance of the MOS observing campaign.  Examples
include galaxy redshift surveys that are designed to measure large
scale structure, such as SDSS \citep{york00}, 2dFGRS
\citep{colless01}, 2SLAQ \citep{cannon2006}, WiggleZ
\citep{drinkwater10}, BOSS \citep{dawson13}, VIPERS
\citep{Scodeggio2016} and 2dFLenS \citep{2dflens}; galaxy surveys that
are designed to measure the detailed physical properties of galaxies,
such as SDSS \citep{york00} and GAMA \citep{gama}; and surveys that
are designed to trace the build-up of stellar mass in the Galaxy, such
as the GALAH survey \citep{DeSilva2015}. 

In the modern era of large surveys, this observing strategy
is being supplemented by ones that incorporate 
{\em dynamic} target selection. In these surveys, target selection is
done shortly before the observation, thus allowing transients to be
observed soon after they are discovered.
This model for MOS surveys is being pioneered by the OzDES survey
\citep{ozdes}, which is obtaining spectra and redshifts for the Dark
Energy Survey \citep[DES;][]{flaugher05}.  Such a strategy will become
standard for future facilities with large, rapidly reconfigurable
fiber positioners, such as 4MOST \citep{4most}, the Subaru Prime Focus
Spectrograph \citep[PFS;][]{sugai12, takada14}, the Dark Energy Spectroscopic
Instrument \citep{DESI2016}, and the Mauna Kea Spectroscopic Explorer
\citep[MSE;][]{mse1, mse2}, all of which will conduct numerous
parallel spectroscopic programs including some related to dynamic
targets discovered by LSST \citep{tyson02, lsstscibook}.

This paper presents results from the first three years of the
OzDES survey, and describes the first OzDES data release.  The
detailed survey description and results from the first year are
presented in \citet{ozdes}.  Briefly, the science objectives of OzDES
include obtaining supernova (SN) host-galaxy redshifts for cosmology
\citep[e.g.,][]{Bazin2011,campbell13},
spectroscopically classifying active transients, monitoring a sample
of active galactic nuclei (AGN) for reverberation mapping \citep[RM --
  see e.g.,][]{bentz09,king15} and potentially for cosmology
\citep[][]{watson11, king14}, securing redshifts for a wide variety of
galaxies to be used for photometric redshift training
\citep[e.g.,][]{sanchez14, bonnett15}, including a large sample of
luminous red galaxies \citep[LRGs;][]{Banerji2015}, and using redshifts of selected galaxies
to confirm their membership in clusters \citep[e.g.,][]{rozo16,
  rykoff16}. OzDES has already produced several discoveries, including
spectroscopy of hundreds of active transients, many new QSOs \citep{tie16}, and the first FeLoBAL QSO in a post-starburst galaxy \citep{Mudd2016}.

Here we focus on the main outcomes from the first three years of
observations, which have comprised just over half of the allocated
observing time for OzDES.  This paper also marks the first public
release of redshifts secured by OzDES, which consists of \ztotal
redshifts from a wide variety of target classes (we note the SN
host-galaxy redshifts will be released in conjunction with future DES
supernova analyses).  We anticipate future data releases from OzDES to
include a final redshift catalogue to be released shortly after the
conclusion of the survey, and an anticipated final release including
the OzDES spectra some time after that.

This paper is organised as follows: Section~\ref{sec:operations}
presents the operational details for the second and third years of
OzDES observations, including a full accounting of fibre allocation.
Section~\ref{sec:redshifts} describes the redshift success rates for
OzDES for various target types as a function of magnitude and spectrum
signal-to-noise.  The first OzDES data release is then described in
Section~\ref{sec:redshift_release}.  In Section~\ref{sec:SN}, we
examine how signal-to-noise depends on exposure time, and in
Section~\ref{sec:future_surveys}, we discuss the prospects for the
full OzDES survey and future multi-object spectroscopy surveys.
Finally, we present concluding remarks in
Section~\ref{sec:conclusions}.
 
\section{OzDES Operations for Years 2 and 3}
\label{sec:operations}

In this Section, we describe several key aspects of the operational
strategy for the second and third years of OzDES observing \citep[the
  analogous information for first year operations was presented
  in][]{ozdes}.  Section~\ref{sec:inst_setup} describes the instrument
setup, calibration strategy, and general observing log.
Section~\ref{sec:data_redux} describes the data reduction procedures,
with a particular emphasis on upgrades from the first year of OzDES
operations.  Finally, Section~\ref{sec:fibre_alloc} presents the
allocation of 2dF fibres to the various OzDES target types.

Throughout this paper we will use the notation Y1/Y2/Y3 to denote the
first/second/third years of OzDES operations, respectively. Y1
corresponds to semester B of 2013 (i.e. August 2013 to January 2014) and includes
data that were taken during a couple of science verification runs at the end semester B in 2012,
Y2 to semester B of 2014, and Y3 to semester B of 2015.  These
correspond to the same operational years for the Dark Energy Survey (DES).

\subsection{Instrument Setup and Calibration Strategy}
\label{sec:inst_setup}

OzDES utilises the AAOmega spectrograph with the 2dF robotic fibre
positioner on the 3.9-metre Anglo-Australian Telescope (AAT) at Siding
Spring Observatory in Australia to target the 10 DES supernova fields,
which cover a total area of 27 sq. degrees. The co-ordinates of the
fields are listed in Table 1 of \citet{ozdes}. 

AAOmega consists of red and blue arms that are split by a dichroic.
We used the x5700 dichroic, which splits the arms at 5700\,\AA. In the blue
arm, we used the 580V grating centred at 4800\,\AA\ in Y1 and Y2, but
in Y3 we shifted the central wavelength by 20\,\AA\ to 4820 \AA\ to
provide larger overlap between the red and blue arms of the
spectrograph. In the red arm, we used the 385R grating centred at
7250\,\AA. This setup provides continuous wavelength coverage from
$\sim 3700$\,\AA\ to $\sim 8800$\,\AA. The overlap between the red and
blue arms is typically $\sim 40$\,\AA\, ($\sim 60$\,\AA\, for Y3),
although this depends on fibre location due to the different spectral
curvature of the red and blue gratings.

The 2dF robotic positioner can place up to 392 science fibres and 8
guide fibres within a 2.1 degree diameter patrol field. The patrol
field of 2dF matches closely the field of view of the DECam imager
\citep{Flaugher2015} on the CTIO 4m Blanco telescope \citep[see Fig.~1
  in][]{ozdes}.

Between Y1 and Y2, the CCDs in the red and blue arms of AAOmega were
replaced.  The new CCD in the blue arm has superior quantum efficiency
and has far fewer cosmetic defects \citep{brough2014}.  Bias and dark frames are used to
remove these defects in the old CCD, but these frames are no longer
needed for the new CCD (though we continue to take a set of dark and
bias frames during each observing run to confirm this).  The new CCD in the
red arm is thicker than the previous one, and has less fringing
at redder wavelengths.  The thicker CCD also has higher quantum
efficiency than the older CCD, especially beyond $\sim 8800$\,\AA, but
is affected more by cosmic rays.

The AAOmega data calibration strategy for Y1 of OzDES is described in
\citet{ozdes}.  Beginning in Y2, we made a number of changes. In
addition to taking flat fields using the flaps that fold in front of
the 2dF corrector (called flap flats), we took flat fields using an
illuminated section of the dome wind-screen (called dome flats). Dome
flats were taken once per run for each
plate\footnote{2dF consists of two plates. While one plate is being
  observed, the other plate is being configured. Once observations for
  a plate are done, the newly configured plate is tumbled into
  position. The average configuration time for OzDES fields is
  40 minutes.}, and are used to correct for the wavelength
dependent throughput differences between fibres that are not corrected by flap
flats. Applying these corrections (often termed
an illumination correction) to the flap flats leads to better
sky subtraction, especially when the sky background is high due to the
presence of the moon. As with Y1, the flap flats are used to determine
the location of the fibres on the CCD and to measure the fibre profiles.

For OzDES Y2 and Y3 observations, we also modified the F-star input
catalogue. F-stars are used to derive sensitivity functions and to
monitor throughput, which mostly depends on seeing and the presence of
clouds. They can also be used to define a zero-point that can be used
to scale spectra from multiple observations before combining
them. Instead of observing stars as faint as $m_r=20$~mag, as we did in
Y1, we restricted the F-stars to the magnitude range $17 < m_r < 18$~mag,
as we found that fainter stars were too faint to reliably determine
the sensitivity functions and to monitor throughput. The magnitude
reported here and throughout the paper is the r-band
magnitude\footnote{We use the {\tt MAG\_APER\_4} r-band magnitude that
  is provided by the DES pipeline \citep{DESDM2012}.}  within a
2\arcsec\ diameter aperture on the DES images. The diameter of the
aperture best matches the diameter of the 2dF fibres projected onto
the sky.

The observing log for Y2 of OzDES is presented in
Table~\ref{tab:observing_log_y2}, and for Y3 in
Table~\ref{tab:observing_log_y3}.  In 2014, OzDES began collaborating
with a new AAT observing program 2dFLenS \citep{2dflens}, which
conducted an extragalactic redshift survey of targets that were more
than 3 magnitudes brighter than the typical OzDES target.  The
shorter exposure time needed by 2dFLenS allowed for better allocation
of short observing windows around the longer OzDES exposures, as well
as more flexibility in scheduling observations at favorable airmasses.  Thus a time
exchange between the two surveys was employed. It also resulted in a
common observing run accounting system (thus yielding non-consecutive
run numbers for OzDES observing runs).


Y2 (Y3) of OzDES was allocated 16 (20) nights on the AAT, of which 5.5
(4.4) nights were lost to weather or instrument failure. A further 3.5 (0.7)
nights were operational but under adverse weather conditions and were
dedicated, in part, to the bright target backup program (see
Section~\ref{sec:fibre_alloc}). The remaining 8.0 (14.9) nights were observed
under ideal conditions and fully dedicated to the main OzDES
program.  Y1 had a total of 14 nights, 4.0 of which were lost to
weather or instrument failure, with the remainder (10.0 nights)
dedicated to the OzDES main program.  
Further OzDES observing beyond Y3 consists of 20
nights each year on the AAT in the second semesters of 2016 (Y4 --
just concluded) and
2017 (Y5), with a final allocation of 12 nights in the second semester
of 2018 (Y6).

\begin{figure}
\begin{center}
\includegraphics[width=0.48\textwidth]{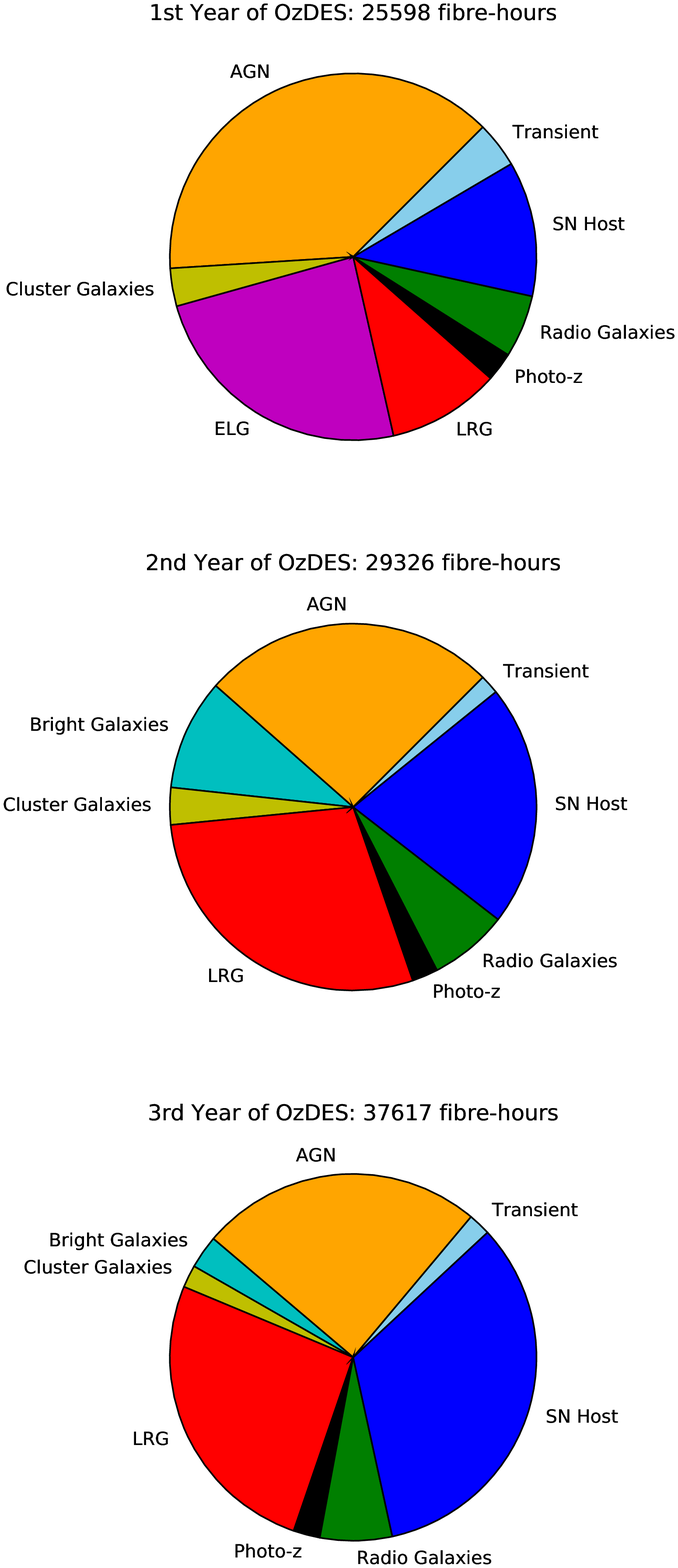}
\caption{Fibre allocation fraction vs. object type for OzDES Y1, Y2, and Y3. Note the steady increase in the number of SN hosts over the first three years.}
\label{fig:fibre_alloc}
\end{center}
\end{figure}

\begin{table*}
  \caption{OzDES fibre allocations by target type for each of the first three years of the survey.}
  \label{tab:fibre_alloc_all}
\begin{tabular}{lrrrrrr}
  \hline
  & \multicolumn{2}{c}{Year 1} & \multicolumn{2}{c}{Year 2} & \multicolumn{2}{c}{Year 3}\\
  Object Type & \multicolumn{1}{c}{Fibre} & \multicolumn{1}{c}{Fraction} & \multicolumn{1}{c}{Fibre} & \multicolumn{1}{c}{Fraction} & \multicolumn{1}{c}{Fibre} & \multicolumn{1}{c}{Fraction} \\
              & \multicolumn{1}{c}{Hours} & \multicolumn{1}{c}{Fibre Time} & \multicolumn{1}{c}{Hours} & \multicolumn{1}{c}{Fibre Time} & \multicolumn{1}{c}{Hours} & \multicolumn{1}{c}{Fibre Time} \\
  \hline
  AGN              &  9827 & 38.3\% &  7602 & 26.0\% &  9373 & 24.9\% \\
  Bright Galaxies  &    28 &  0.1\% &  2863 &  9.8\% &  1100 &  2.9\% \\
  Cluster Galaxies &   854 &  3.3\% &   956 &  3.3\% &   747 &  2.0\% \\
  ELGs             &  6175 & 24.1\% &    -- &     -- &    -- &     -- \\
  LRGs             &  2539 &  9.9\% &  8400 & 28.6\% &  9788 & 26.0\% \\
  Photo-z Targets  &   649 &  2.5\% &   668 &  2.3\% &   883 &  2.3\% \\
  Radio Galaxies   &  1409 &  5.5\% &  2032 &  7.0\% &  2386 &  6.3\% \\
  Strong Lens      &    45 &  0.1\% &    81 &  0.3\% &    -- &     -- \\
  SN Hosts         &  3041 & 11.9\% &  6230 & 21.3\% & 12600 & 33.5\% \\
  Live Transients  &  1031 &  4.0\% &   494 &  1.7\% &   740 &  2.0\% \\
  {\bf Total}      & {\bf 25598} &  & {\bf 29326} &  & {\bf 37617} &  \\
  \hline
\end{tabular}
\end{table*}

\subsection{Data Reduction}
\label{sec:data_redux}

The processing of the raw data is described in detail in
\cite{ozdes}. Here we briefly summarize the main steps. Each
observation consists of one or more exposures of the field, an arc
frame, and two fibre flats. To get sufficient counts in the fibre
flats for both arms, one requires two exposures with different
exposure times. The fibre flats are used to locate and trace the
spectra on the CCDs (the so-called tram line maps), and the arcs are
used to set the wavelength scale.

For each exposure, the overscan region is used to remove the bias, and
for data taken with the old blue CCD in Y1 only, bias and dark
frames are used to reduce the impact of cosmetic defects (see
Sec.~\ref{sec:inst_setup}). The one-dimensional spectra are then
extracted using the tram line maps to guide the extraction and then
resampled to a common wavelength scale.

An important processing step is the removal of the coherent residuals
that remain after subtracting a spectrum of the night sky\footnote{The
  spectrum of the night sky is determined from the 25 sky fibres that
  are distributed across the 2dF field-of-view.}  \citep{sharp10}.
Without this step, the residuals impede the identification of real
spectral features.

We used a modified copy of version 6.2 of \tdfdr\ \citep{croom04} to
process the data from AAOmega. Both the data from Y1, which had used
an earlier version of \tdfdr, and the data from Y2+Y3 have been
processed consistently with the newer version.

The publicly available version of \tdfdr\ improves on the earlier version of
\tdfdr\ used in \citet{ozdes} in a number of ways. Tram line maps
(the location of the spectra on the detectors) are more precise,
and one can now simultaneously fit a model of the background scattered
light together with the flux in the fibres for each column of the
detector. The scattered light is modeled with splines and the fibre
profiles are modeled with Gaussians. The widths of the Gaussians and
their locations are determined from the flat fields.

The publicly available version of \tdfdr\ was modified in several key areas:

\begin{itemize}
\item Using PyCosmic \citep{pycosmic} instead of LACosmic \citep{vanDokkum2001}
to locate pixels affected by
  cosmic rays. These pixels are marked as bad and are removed from
  further analysis.
\item Using singular value decomposition (SVD) when the least-squares
  solution fails to produce an adequate solution to the matrix
  equation that is used to optimally extract the flux in the fibres
  \citep[see][for details]{sharpbirchall10}. On average, we use SVD instead of
  least squares once every 200 columns. Without this modification, the data in these columns
  would have been treated as bad and not used.
\item Average over 10 columns when fitting the fibre profile for low
  signal-to-noise regions in the flat fields. This is most severe in
  the blue end of the spectral range. This results in more accurate
  measurement of the fibre profile, which leads to more accurate
  extraction of the flux from the spectra.
\item Applying an illumination correction to the flap flats using the
  dome flats, which, as noted above, leads to better sky subtraction.
\item Applying a different set of sensitivity functions for the
  instrument before and after the upgrade of the CCDs.
\end{itemize}

Once the data from the red and blue arms have been processed with
\tdfdr, we merge the red and blue spectra into a single
spectrum. Since some targets are observed over multiple nights and can
appear in more than one field (because of field overlap) we sum all
the data on a single target into one spectrum, first scaling each
spectrum (the variance is scaled by the square of this scaling) and
then weighting by the inverse of the variance in the sum. We scale the
spectrum with either the inverse of the median value of the spectrum
or the inverse of 0.1 times the square root of the median of the
variance, whichever turns out to be the smallest. This accounts for
cases in which the median flux is close to zero, which would result in
very large scale factors. The 0.1 factor was chosen after some
experimentation. We plan to revisit this choice for the next data
release.

While the processing has improved, there is still room for further
improvement. In particular, one can sometimes see a discontinuity in
the spectra near the wavelength where the dichroic splits the two arms
of the spectrograph. This is largely due to errors in sky subtraction
and is caused by a combination of a residual background (sometimes caused by the wings of a bright star\footnote{Usually
  this was unintended. In regions close to stars that are heavily
  saturated in the DES images, source detection algorithms, such as
  SExtractor, can erroneously break the image of the star into many
  false objects. Some of these false objects make it into the source
  catalogues. The problem mostly affects data from Y1. Greater scrutiny of the input catalogues during Y2 and Y3 largely
eliminated this problem.}), and poor spectral uniformity in the flap flats
(mostly affects Y1 data when dome flats were not
taken). The accuracy of the sky subtraction, which is assessed by
examining how well the continuum of the sky is removed from the sky
fibres, is about 1\% in the best cases and 6\% in the worst cases.

We are investigating techniques to further improve the processing. In
addition to removing the spectral inhomogeneity in the Y1 flap flats and
improving the modeling of the background, we are experimenting with more
complex fibre profiles. In detail, the fibre profile is not
Gaussian. The core of the profile is more boxy than a Gaussian and there
are broad exponential wings. We anticipate that these improvements will be
available before the end of OzDES.

\subsection{OzDES Target Allocations for Y2 and Y3}
\label{sec:fibre_alloc}

The target selection procedure and primary target classes for OzDES
were described in detail in Section 3.2 and 3.3 of \citet{ozdes},
respectively.  A number of changes, which are described in detail
below, were implemented for Y2+Y3.  Fibre observing time allocations
for Y1+Y2+Y3 of OzDES are summarised in
Table~\ref{tab:fibre_alloc_all}, and presented in
Figure~\ref{fig:fibre_alloc}.

The OzDES active galactic nuclei (AGN) reverberation mapping (RM)
project \citep{king15} transitioned from Y1 operations of
identifying AGN suitable for RM followup to Y2 operations of actively
monitoring AGN selected in Y1.  This decreased the fraction of
fibres allocated to AGN, as the Y1 target selection program had
concluded.  OzDES now has a core sample of 771 AGN that are monitored for
the RM program across all 10 DES-SN fields, which should consistently
account for 25\% of fibre allocations for the remainder of the
project.  

The allocation of fibres to supernova host-galaxies\footnote{This includes the hosts of transients that may not be supernovae.} nearly doubled
from Y1 to Y2, and increased a comparable amount between Y2 and Y3.
This is a consequence of both the increase in number of confident
prior SN candidates from DES and an extension of the SN host magnitude
limit to include fainter hosts (down to $m_r=24.0$~mag).  The
increase of DES Y1 SN candidates is partly due to an improved
reprocessing of the DES Y1 imaging data \citep{kessler15} yielding
several hundred new SN candidates with hosts to be targeted by OzDES.

In the first year of OzDES observing, we found that adverse observing
conditions (i.e. poor seeing or significant extinction) resulted in
low redshift success rates (of order 10\%).  Thus at the end of Y1 we
implemented a poor weather backup program comprised primarily of a
new ``bright galaxy'' target class, consisting of galaxies with
integrated magnitude of $18 \lesssim m_r \lesssim 19.5$~mag identified in
DES imaging but not tied to an existing DES spectroscopic followup
program.  The goal of this program is to maximise the scientific
utility of data taken with OzDES even in adverse conditions by
expanding the extragalactic redshift catalogues in the DES SN fields,
many of which have significant archival imaging data across the
electromagnetic spectrum.  Such a program will help with photo-z
calibration and SN lensing measurements.

The bright program was triggered at the discretion of the observers, 
and was typically triggered when the seeing was worse than 3\arcsec, or there
was significant cloud cover.
An important consideration in making the decision to switch to the backup program is the
40-minute configuration time for a 2dF plate, meaning the decision to
switch to the bright program would have at least a 40-minute
delay before implementation.

OzDES target selection procedures were modified for emission line
galaxies (ELGs) and luminous red galaxies (LRGs) in Y2 and Y3. After
Y1, it became apparent that we would not be able to obtain 10,000
redshifts for ELGs and 10,000 redshifts for LRGs with the number of
spare fibres available during the survey. Therefore, after Y1, ELGs
were no longer targeted for photometric redshift training, releasing a
large fraction of fibres.  Accordingly, the fraction of fibres
allocated to LRGs increased, as these LRGs were of equal priority in
the target selection algorithm to the ELGs that were removed
\citep[see Section 3.2 of][for details]{ozdes}. The follow-up of ELGs
in the DES fields was continued at other facilities.

After Y1, OzDES also began targeting galaxies selected with the redMaGiC
\citep{rozo16} algorithm, which selects LRG galaxies with accurate and
unbiased photometric redshifts. These galaxies are being used in large
scale structure and weak lensing studies in DES.  RedMaGiC galaxies are counted
as LRGs in Table \ref{tab:fibre_alloc_all} and Fig.~\ref{fig:fibre_alloc}. We also targeted Brightest
Cluster Galaxies (BCGs) from the SpARCS catalogue \citep{webb15}, tertiary
calibration stars in the DES supernova fields, host-galaxies of exotic
transients from SNLS \citep[following][]{lidman13}, and faint QSOs in
the S1 and S2 fields (those which overlap with SDSS Stripe 82).

Finally, the sample of radio-galaxies in Y2 and Y3 was extended to
galaxies with lower radio luminosities. Many of these galaxies tend to
be relatively bright nearby star-forming galaxies. This leads to
brighter galaxies being targeted in Y2+Y3 than in Y1, as can be seen
in the bottom left hand plot of
Figure~\ref{fig:redshift_completeness}.

\section{Redshift Results from the first three years of OzDES}
\label{sec:redshifts}

\subsection{Redshift Outcomes}
\label{sec:redshift_outcomes}

For Y1 and Y2, redshifts were obtained using the {\tt RUNZ}
redshifting software, which was developed by Will Sutherland for the
2dF Galaxy Redshift Survey. In Y3, we switched to {\tt
  Marz}\footnote{Manual and automatic redshifting software,
http://samreay.github.io/Marz/}
\citep{Hinton2016}, after first verifying that the redshifts from {\tt
  RUNZ} and {\tt Marz} were the same. {\tt Marz} is an open-source,
client-based, web-application that provides most of the
functionalities of {\tt RUNZ} as well as new features, but none of the
difficulties associated with installing and compiling {\tt RUNZ}.

As for Y1, each source was inspected by two redshifters. A third
redshifter inspected their work to resolve conflicting
assignments and merge the results.

At the conclusion of Y3, OzDES has obtained spectra for a total of
\targets targets yielding \ztotalAll successful redshifts (a
successful redshift has $Q \ge 3$, see
Section~\ref{sec:redshift_conditions}).
Table~\ref{tab:redshift_counts} presents the number of redshifts for
each target type at the end of each of the first three years of OzDES.
We note that some targets are triggered for spectroscopy by multiple
DES science working groups.  Over multiple runs, an object that was
tagged with a certain type (e.g. Photo-z or ELG) in an earlier run may
have been tagged in a later run with another type (e.g. SN host) and
reobserved. This is one reason why the number of objects with
redshifts can drop from one year to the next, as is the case for ELGs
between Y2 and Y3. The number can also change if the data were
reprocessed and re-analysed between the OzDES observing seasons.

The number of observing nights increased progressively from Y1 to Y3
(as did the fibre-hours on sky, see Table~\ref{tab:fibre_alloc_all}),
but the redshift success rate slowed over this period.  The change in
the redshift success rate is due to an increase in the relative number of
fainter objects being targeted, as brighter targets require less
exposure time to obtain redshifts and are thus removed from the OzDES
target pool earlier.  

\begin{figure*}[h]
\begin{center}
\includegraphics[width=0.96\textwidth]{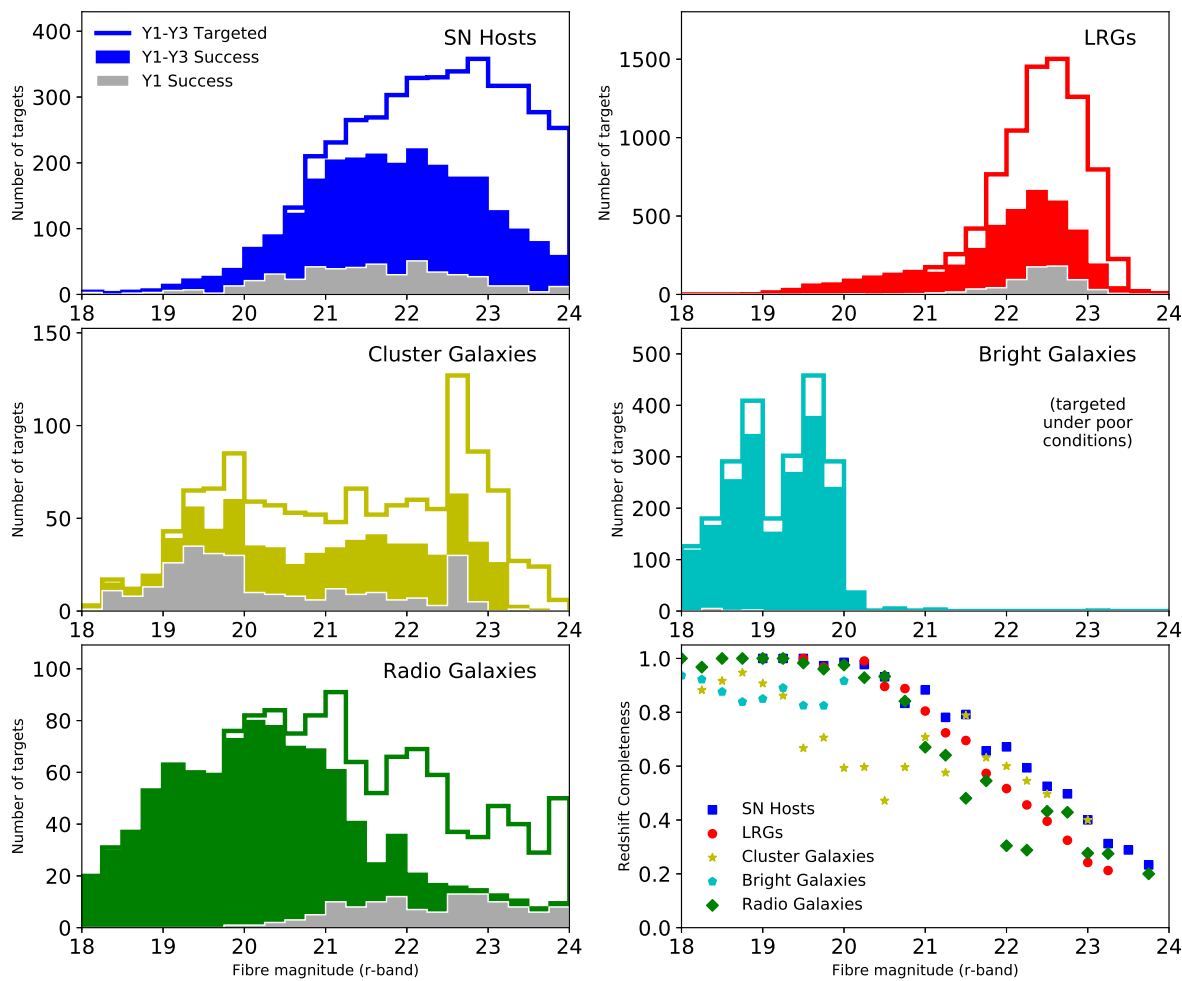}
\caption{Fibre magnitude distribution of all objects targeted by OzDES from Y1-Y3 (white histogram), objects with successful redshifts at the end of Y3 (filled histogram), and (for comparison) those with successful redshifts at the end of Y1 \citet[as in Figure 4 from][]{ozdes}.  Histograms for five categories of objects are shown, and in the lower right panel we show the redshift success fraction (number of successful targets divided by full number of targets in each magnitude bin) as a function of magnitude for the five target classes.}
\label{fig:redshift_completeness}
\end{center}
\end{figure*}

Figure~\ref{fig:redshift_completeness} presents the magnitude
distributions of several target classes, separated by redshift status:
all targeted objects (open histograms), all objects with secure
redshifts for Y1-Y3 combined (filled coloured histograms), and those
with redshifts obtained in Y1 of OzDES (filled grey histograms).  We
also show the redshift completeness versus r-band magnitude for these
target classes.  

\begin{center}
\begin{table}
  \caption{OzDES cumulative redshift counts by target type for the first three years of the survey. Not all target types are listed.}
  \label{tab:redshift_counts}
 \begin{tabular}{lrrr}
  \hline
  & \multicolumn{1}{c}{End Y1} & \multicolumn{1}{c}{End Y2} & \multicolumn{1}{c}{End Y3}\\
  Object Type & \multicolumn{1}{c}{\# Redshifts} & \multicolumn{1}{c}{\# Redshifts} & \multicolumn{1}{c}{\# Redshifts}  \\
  \hline
  AGN              &  1781 &   1901 & 1916 \\
  Bright Galaxies  &    24 &   2183 & 2902 \\
  Cluster Galaxies &   284 &    525 &  732 \\
  ELGs             &  1293 &   1429 & 1419 \\
  LRGs             &   711 &   2528 & 4108 \\
  Photo-z Targets  &  1622 &   1741 & 1740 \\
  Radio Galaxies   &   143 &    788 & 1027 \\
  Strong Lenses    &     3 &     14 &   14 \\
  SN Hosts         &   498 &   1345 & 2566 \\
  Live Transients  &   232 &    369 &  494 \\
  Other            &    47 &    266 & 1658 \\
  {\bf Total}      & {\bf 6648} & {\bf 13089} & {\bf 17893}  \\
  \hline
\end{tabular}
\end{table}
\end{center}

\begin{figure*}
\begin{center}
\includegraphics[width=0.96\textwidth]{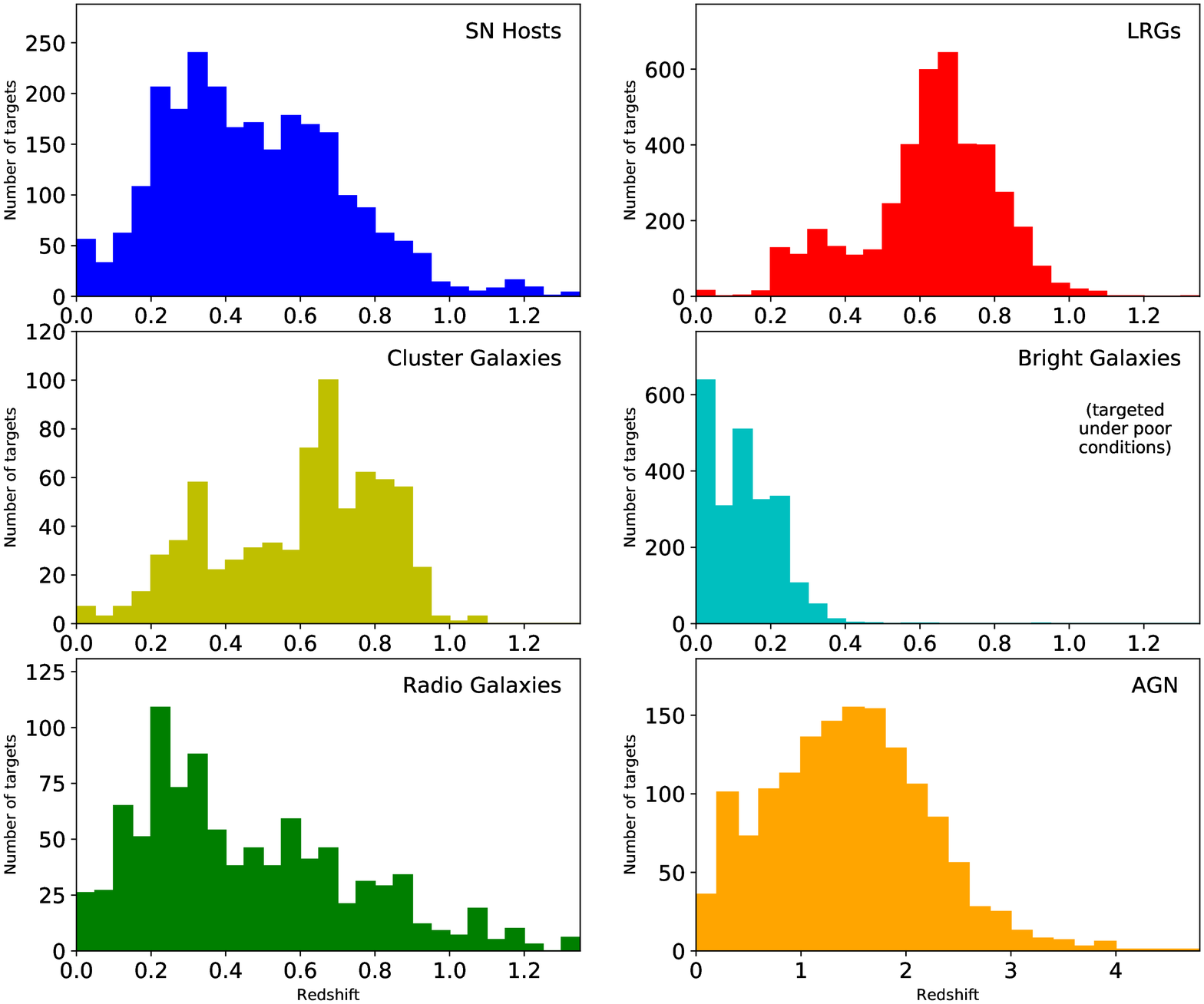}
\caption{Redshift distributions at the end of Y3 for the 5 different 
object types shown in Fig.~\ref{fig:redshift_completeness}. Also shown, in the lower right panel, is the redshift distribution for AGN. Note that the redshift scale of the AGN plot differs from the other plots.}
\label{fig:redshift_completenessdistribution}
\end{center}
\end{figure*}

The redshift success rate shows a characteristic trend with magnitude
for most target types, where we have a high likelihood of obtaining a
redshift for targets brighter than $m_r=21$~mag in a single observation.
As we will show below, this is driven by signal-to-noise.  Fainter
objects, which continue to be targeted, are steadily accumulating the
necessary signal-to-noise to obtain a redshift.

In Fig.~\ref{fig:redshift_completenessdistribution}, we show redshift
distributions for the five object classes shown in
Fig.~\ref{fig:redshift_completeness} as well as for AGN.  The SN hosts
extend up to $z\sim1.2$. At this redshift the [OII] $\lambda$3727 doublet
is close to the red end of the spectral range covered by OzDES
spectra. In principle, one could push to higher redshifts by changing
the wavelength settings of AAOmega, but this comes at the expense of
losing coverage in the blue, which would have an impact in the number
of redshifts that are obtained for objects at lower redshifts.  The
AGN sample is intrinsically brighter and thus extends to much higher
redshifts -- this sample consist of objects that are part of the AGN
reverberation mapping programme, additional AGN that were screened during
Y1, and faint QSOs that are observed in the S1 and S2 fields.

\subsection{Conditions for Redshift Success}
\label{sec:redshift_conditions}

We examine the likelihood of securing a redshift, first as a function of
signal-to-noise ratio, and then as a function of magnitude. In
Sec.~\ref{sec:future_surveys}, we'll use these likelihoods to compute
the likely yield of SN host redshifts by the end of the survey.

OzDES assigns a redshift quality flag to each redshift. These values are
described in detail in \citet{ozdes}, but are summarised below:

\begin{itemize}
  \item $Q=4$, redshift based on multiple strong spectroscopic features matched, $>99$\% confidence.
  \item $Q=3$, redshift based typically on a single strong spectroscopic feature or multiple weak features, 95\% confidence.
  \item $Q=2$, potential redshift associated with typically a single weak feature, low confidence.
  \item $Q=1$, no matching features, thus no constraints on redshift.
  \item $Q=6$, securely classified star.
\end{itemize}

\begin{figure*}
\begin{center}
\includegraphics[width=0.245\textwidth]{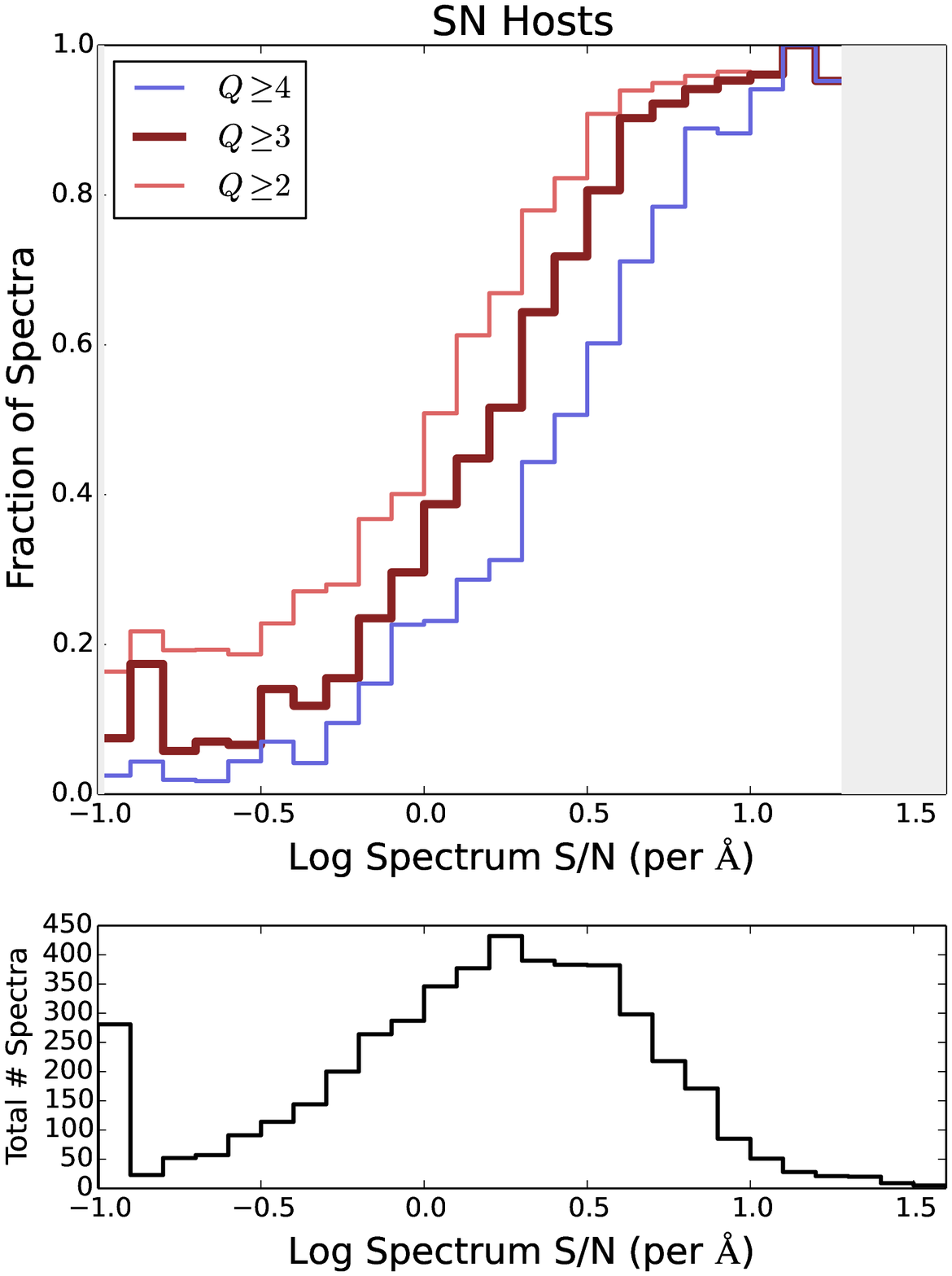}
\includegraphics[width=0.245\textwidth]{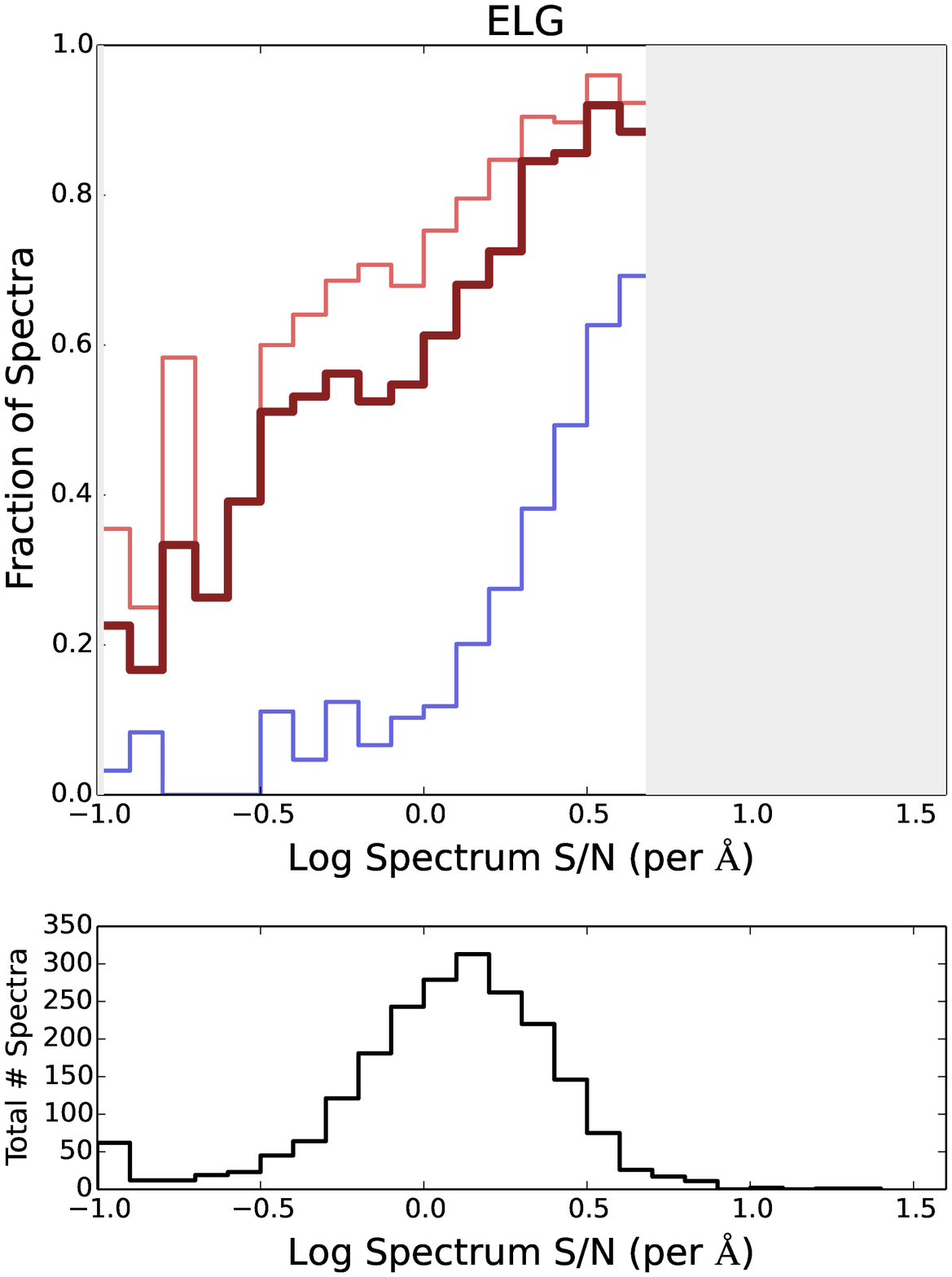}
\includegraphics[width=0.245\textwidth]{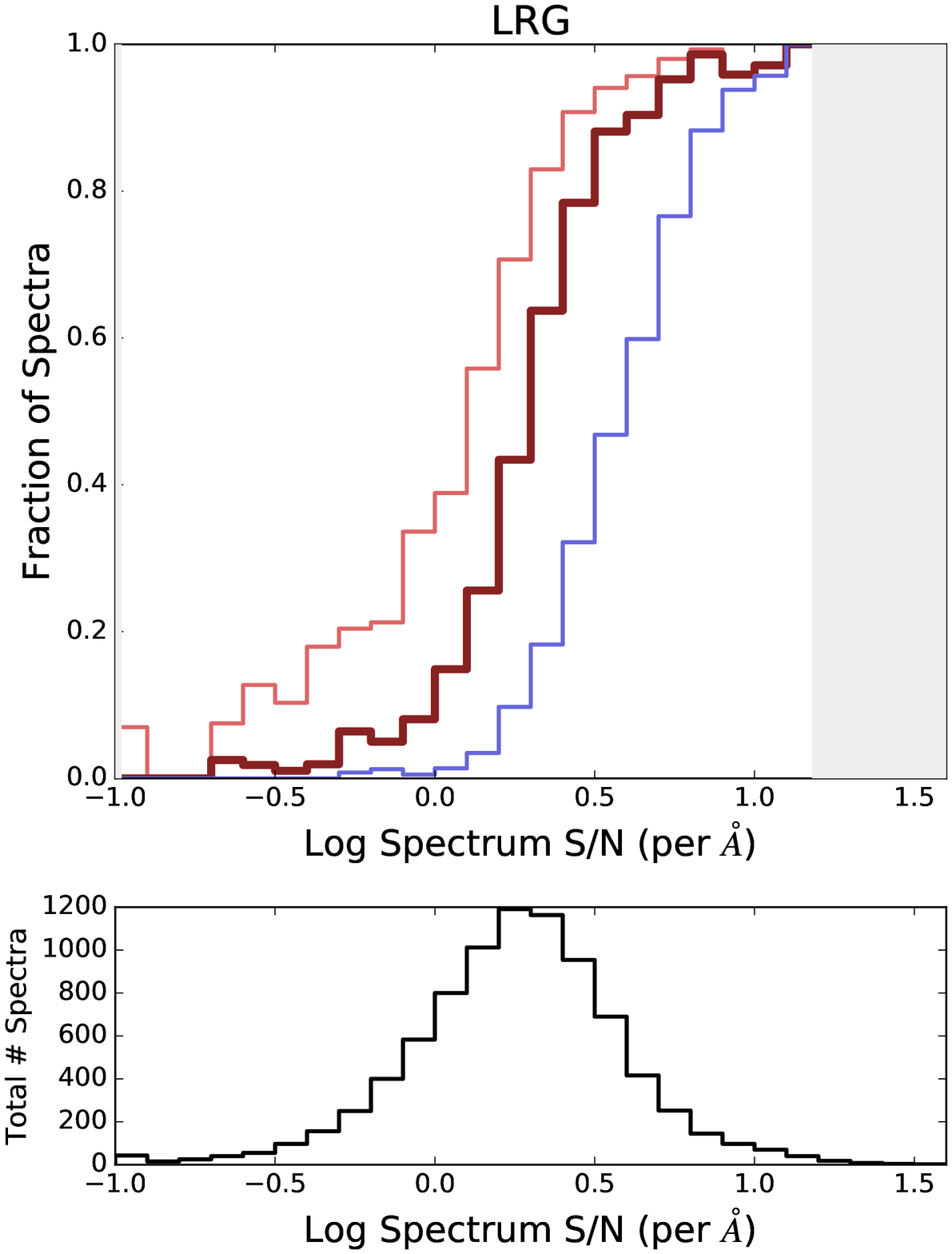}
\includegraphics[width=0.245\textwidth]{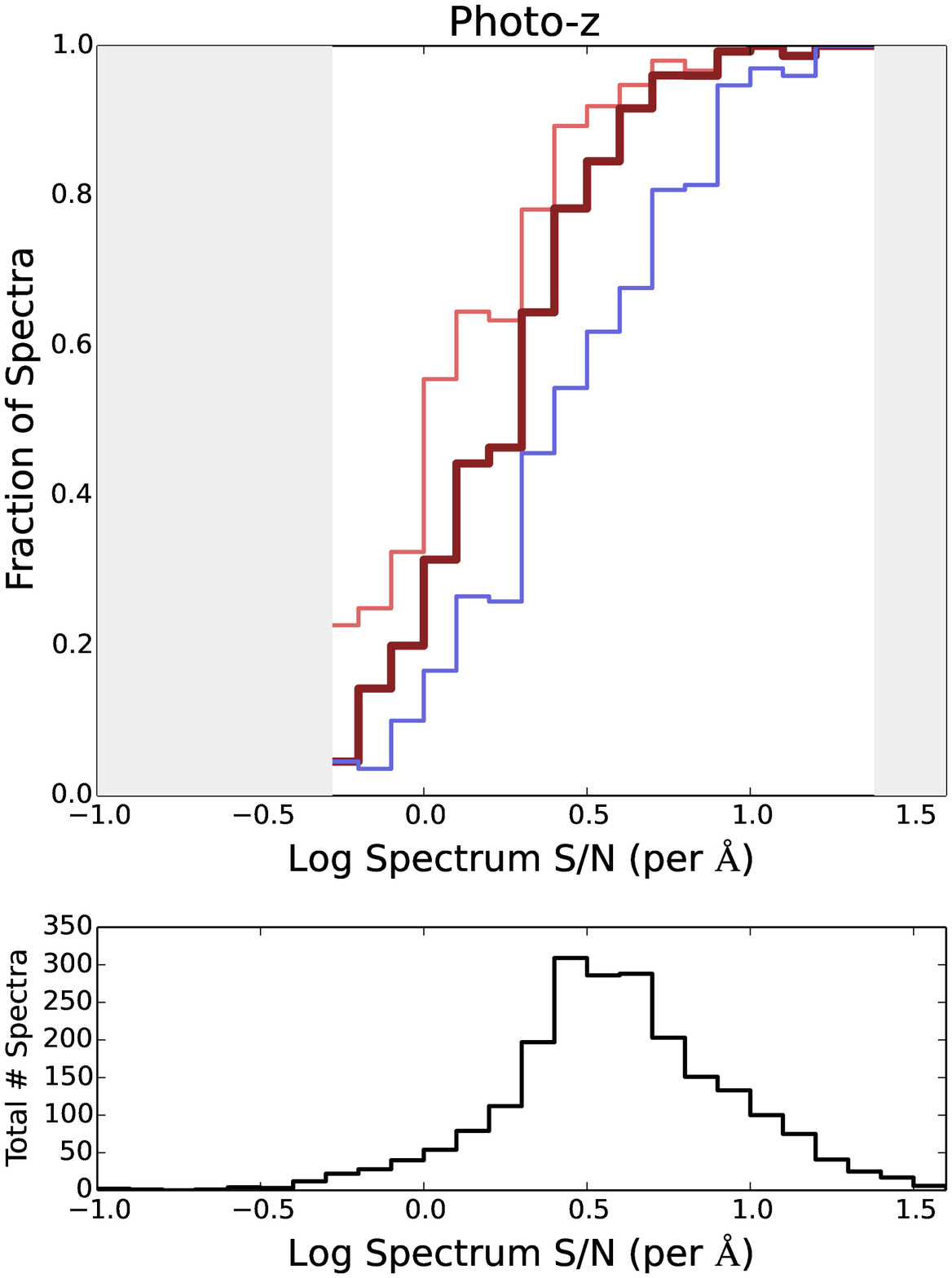}
\caption{Distributions of spectra signal-to-noise (S/N) for the final redshift quality flag ($Q$) groups for each type of OzDES target.  The top panels show the fraction of targets achieving a quality flag $Q$ as a function of S/N for several target types, while the bottom panels show histograms of the S/N for those same target types.  In the upper panels bins with less than 20 total spectra are omitted (denoted by grey shaded areas).}
\label{fig:qop_stats}
\end{center}
\end{figure*}

In Figure~\ref{fig:qop_stats}, we plot the fraction of targets achieving a given redshift quality $Q$ as a function of the (logarithmic) S/N for spectra.  This is plotted for selected object classes, split by
the three
(extragalactic) $Q$ values.  These are calculated as the mean S/N per
\AA\ in the wavelength range 6500 \AA\ $\leq \lambda \leq$ 8500 \AA,
which is redward of the 4000 \AA\ break for intermediate redshift ($z
\sim 0.5$) galaxies, and is where the throughput of AAOmega is
greatest (thus this wavelength range typically dominates whether or
not a successful redshift is achieved).

These redshift quality plots show that for most target categories
(including those not shown here), the S/N at which the
likelihood of securing a redshift ($Q \geq 3$) exceeds 50\% is
2 to 3 in 1-\AA\ bins (i.e. $\log(\mathrm{S/N}) \approx 0.3$). 
The one obvious exception to this trend is ELGs (second column of
Figure~\ref{fig:qop_stats}) which have a flatter trend of success
versus spectrum S/N -- this is because the average spectrum S/N is
weakly correlated with the strength of the narrow oxygen doublet
responsible for most successful ELG redshifts.

\begin{figure}
\begin{center}
\includegraphics[width=0.50\textwidth]{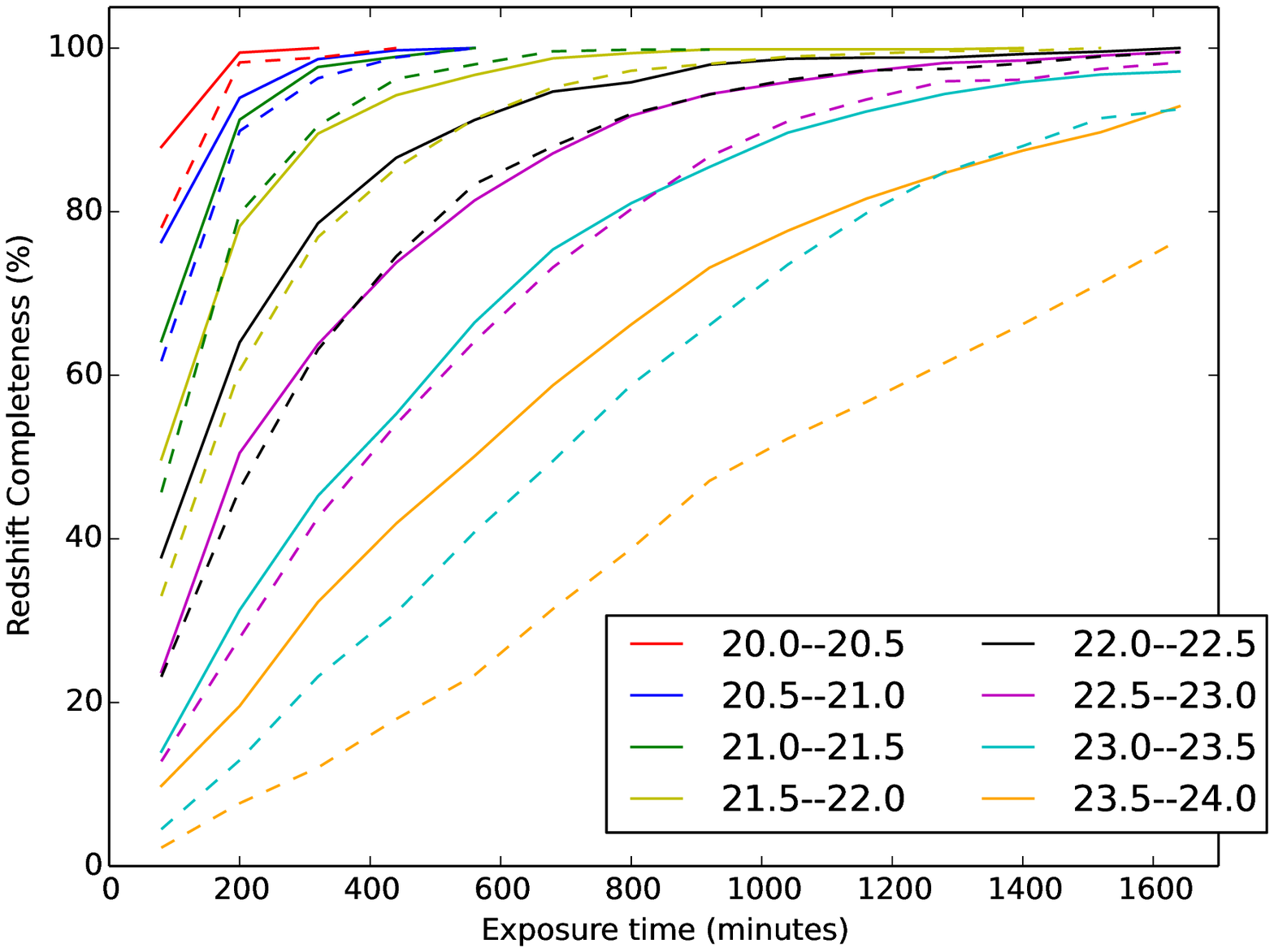}
\caption{Redshift completeness versus exposure time for different magnitude bins in the SN host sample.  For each magnitude bin, a solid line is shown to present all successes ($Q\ge3$) and a dashed line of the same colour is shown to present the most secure redshifts ($Q\ge4$). Note how the fraction of sources with a redshift monotonically increases with exposure time for all objects, irrespective of magnitude.}
\label{fig:completeness_mag}
\end{center}
\end{figure}

In Fig.~\ref{fig:completeness_mag}, we examine how the redshift
completeness changes as a function of r-band magnitude.  We limit
the analysis to SN hosts, as they cover a broad range of magnitudes
and galaxy types, and represent the largest source of uniformly
selected targets. As expected, redshift completeness increases with
exposure time. We reach a completeness of over 90\% for objects down to
$m_r=23.75$~mag, which is the midpoint of the last magnitude bin in
Fig.~\ref{fig:completeness_mag}. We use
Fig.~\ref{fig:completeness_mag} to estimate the yield of SN host
galaxies that we expect by the end of Y6 in Sec.~\ref{sec:future_surveys}.

\subsection{Redshift Reliability}
\label{sec:redshift_reliability}

We now turn the question of how reliable the OzDES redshifts are,
particularly for those with a redshift quality flag $Q=3$, which are
nominally expected to be correct 95\% of the time.  Fortunately, for
the SN host galaxies in the OzDES sample, we did not remove these
objects from the target queue once they were marked as a $Q=3$
redshift, but only once a highly secure $Q=4$ redshift was obtained.

\begin{figure}
\begin{center}
\includegraphics[width=0.48\textwidth]{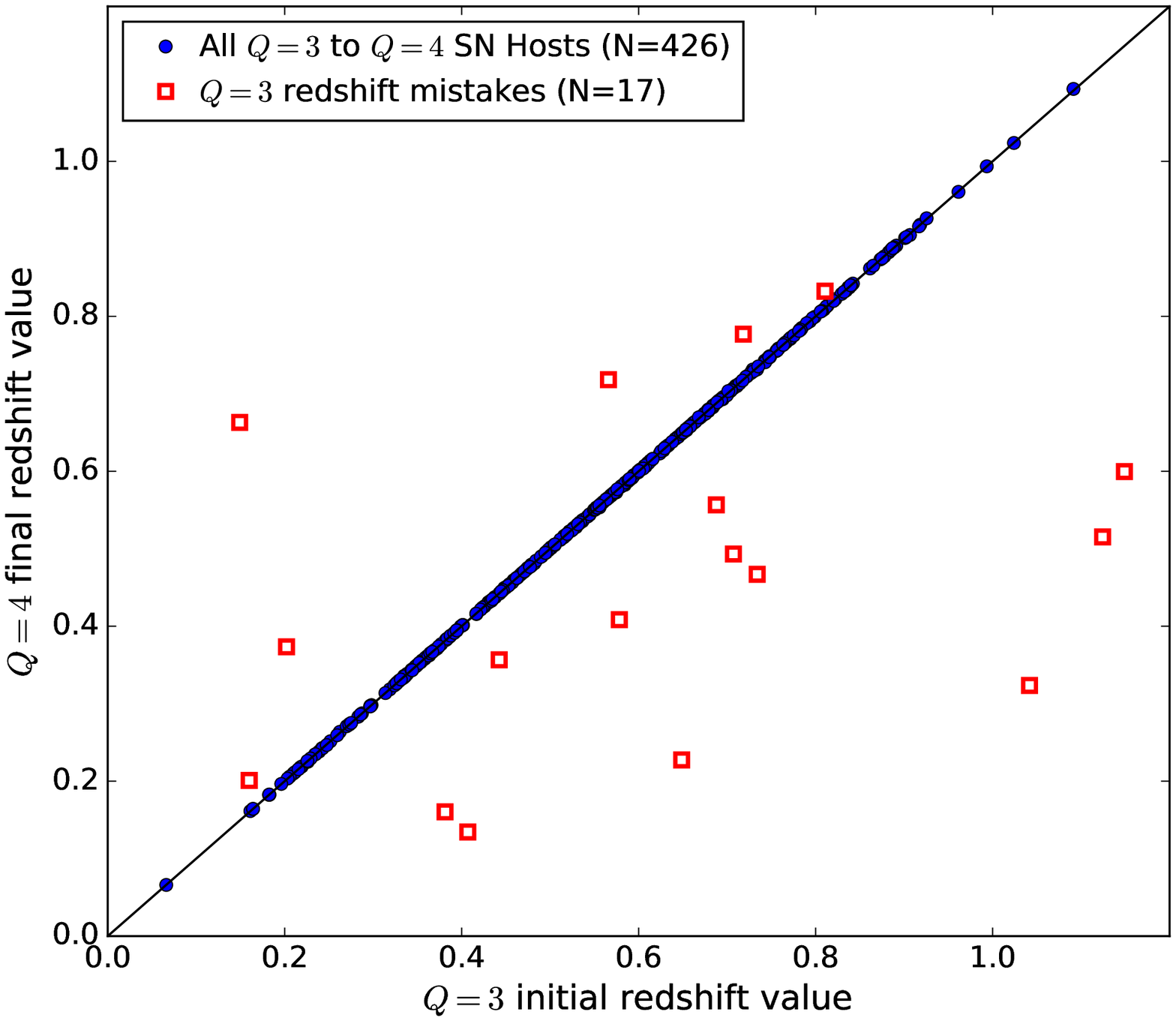}
\caption{Redshift comparison for OzDES SN hosts with initial redshift quality $Q=3$ that were successfully re-observed until attaining $Q=4$.  The full sample is shown as small blue dots ($N=426$ objects) while the ($N=17$) outliers (defined as $\Delta z > 0.003$) are shown as large red squares.}
\label{fig:redshift_reliability}
\end{center}
\end{figure}

We thus examined the redshift catalogs compiled at the end of each
observing run to examine which SN hosts had initially achieved a
successful redshift with $Q=3$ but ultimately succeeded in achieving a
$Q=4$ redshift.  Out of the \zhost SN hosts with successful redshifts,
426 have a final quality flag of $Q=4$ but previously had a $Q=3$
quality flag at the end of a prior run.  In
Figure~\ref{fig:redshift_reliability} we plot the final ($Q=4$)
redshift against the initial ($Q=3$) redshift for these SN hosts.

We define outliers as those with a redshift change of
$\Delta z > 0.003$ (note we obtain the same sample if we define this
as $\Delta z/(1+z) > 0.003$).  This provides 17 objects (out of 426)
whose initial $Q=3$ redshift was not consistent with its final $Q=4$
redshift.  This yields a redshift error rate of 4\% for $Q=3$
redshifts.  Implicit in this calculation is the assumption of a zero
redshift error rate for $Q=4$ redshifts. Using objects in the overlap
regions between DES fields that have been observed more than once,
\cite{ozdes} found no inconsistent $Q=4$ redshifts out of 136
objects. Hence, using a zero redshift error rate for objects with
$Q=4$ to derive a redshift error rate of 4\% for $Q=3$ redshifts is
reasonable. It is similar to the error rate for $Q=3$ redshifts found
in \citet{ozdes}.  We also note the RMS difference between the  $Q=3$
and $Q=4$ redshift values is $\sigma_z = 4.2 \times 10^{-4}$, consistent
with the galaxy redshift uncertainty reported in \citep{ozdes}.

The number of SN hosts with an erroneous redshift will be smaller than
4\%, since we only remove SN hosts from the observing queue objects
once $Q=4$. The ratio of the number of SN hosts with a $Q=3$ redshift
to the number of SN hosts with a $Q=4$ redshift is 1 to 4. Hence the
number of SN hosts with an erroneous redshift will be less than 1\%.
Note this value for redshift reliability is related to redshifts secured
for the nominal SN host.  The additional error of incorrectly identified
hosts \citep[shown in][to be around 3\%]{Gupta2016} is important for
\snia\ cosmological analyses, but it beyond what we examine here.

\section{The First OzDES Redshift Release}
\label{sec:redshift_release}

This paper marks the first release of data from the OzDES survey,
which we henceforth refer to as OzDES-DR1\footnote{The redshift
  catalogue will be made available from the Strasbourg Astronomical
  Data Center and http://www.mso.anu.edu.au/ozdes/DR1}. OzDES-DR1
consists of the redshift catalogue from the first three years of
observing, and includes coordinates and redshifts for all target
categories in the OzDES program except supernova host galaxies.  The
redshifts for SN hosts will be published after the OzDES survey
ends. OzDES-DR1 includes confident extragalactic redshifts ($Q=3$ and
$Q=4$), as well as confirmed foreground Milky Way stars ($Q=6$).  In
total, this first redshift catalogue contains \ztotal redshifts in the
ten DES supernova fields.

\begin{figure}
\begin{center}
\includegraphics[width=0.48\textwidth]{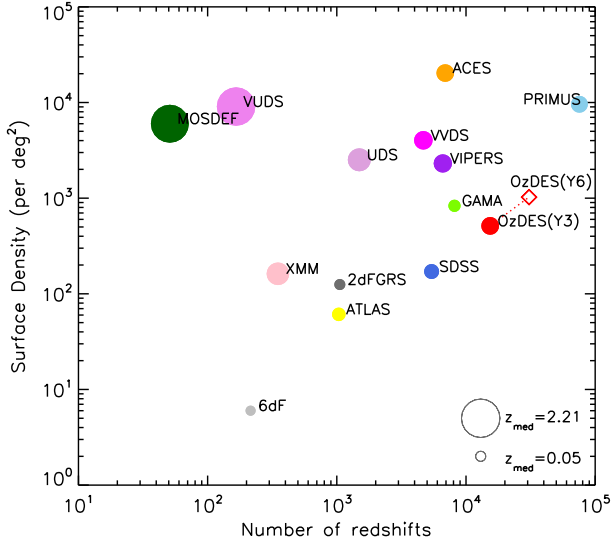}
\caption{Comparison of OzDES-DR1 against outcomes for other major
  redshift surveys in the OzDES fields.  The average redshift surface
  density of the survey is plotted against total number of redshifts
  in the OzDES fields, with the size of each point corresponding to
  the median redshift of the survey in our fields.  OzDES-DR1
  represents the second highest number of redshifts produced in these
  fields, with a surface density approaching that of surveys on larger
  (8m-class) telescopes. As time progresses, this point will move
  further up and to the right, as shown by the dotted line.}
\label{fig:redshift_surveys}
\end{center}
\end{figure}

OzDES-DR1 also includes the primary target type for each object.
We note several important caveats associated with this information.
First, some objects were selected for spectroscopic followup by
multiple science working groups within the DES collaboration.  For
practical purposes, OzDES tagged each multiply-requested object by the
target category with highest (active) priority for followup \citep[see
  Section 3.2 and Table 3 of][for further details on the target
  categories and priorities]{ozdes}.  Second, we caution against
interpreting the set of objects in a given category as a complete
demographic sample.  There are two main reasons for this: i) the
selection criteria for some target types evolved during the course of
the first three years, and ii) objects that were initially included early in the
OzDES target list might have been deselected once a redshift became
available from a recently published survey (e.g.~GAMA or VIPERS).


Objects identified from DES imaging as targets for spectroscopy were
first checked against a compilation of redshifts in the DES supernova
fields from surveys listed in Table~\ref{tab:grc_inputs}, which also
lists the target density of these surveys and the median redshift for
their targets in the OzDES fields.  Each DES science working group
would apply different criteria to de-select objects.
We note that OzDES has a
higher on-sky target density than many wide-area surveys (such as 6dF,
2dFGRS, SDSS) and a similar redshift range to many deep surveys (such
as ACES, PRIMUS, VIPERS, VVDS -- see
Figure~\ref{fig:redshift_surveys}).

\begin{table}
\centering
\caption{Extragalactic redshift catalogues searched for existing redshifts for DES photometric targets sent to OzDES.}
\label{tab:grc_inputs}
\begin{tabular}{lrrrc}
\hline
Catalog & \# Redshifts in & $z_\mathrm{med}$~$^a$ & $\Sigma_z$~$^b$ & Ref. \\
        & OzDES fields    &            & [deg$^{-2}$]    &  \\
\hline
2dFGRS       &  1055 & 0.12 &   125 &  1 \\
6dF          &   215 & 0.05 &     6 &  2 \\
ACES         &  6908 & 0.62 & 20320 &  3 \\
ATLAS        &  1036 & 0.31 &    61 &  4 \\
GAMA         &  8143 & 0.21 &   832 &  5 \\
MOSDEF       &    51 & 2.17 &  6000 &  6 \\
PRIMUS       & 75511 & 0.56 &  9583 &  7 \\
SDSS         &  5417 & 0.42 &   171 &  8 \\
UDS          &  1493 & 1.03 &  2518 &  9 \\
VIPERS       &  6615 & 0.67 &  2307 & 10 \\
VUDS         &   166 & 2.21 &  9075 & 11 \\
VVDS         &  4676 & 0.69 &  4025 & 12 \\
XMM          &   350 & 1.00 &   162 & 13 \\
{\bf OzDES}  & 15388 & 0.63 &   513 &  \\
\hline
\end{tabular}
\\
\flushleft
$^a$ Median redshift of the surveys' redshifts in the OzDES fields. \\
$^b$ On-sky redshift density of the (entire) survey. \\
{\bf References:}
(1) \citet{colless01};
(2) \citet{jones09};
(3) \citet{cooper12};
(4) \citet{mao10, mao12};
(5) \citet{liske15};
(6) \citet{kriek15};
(7) \citet{coil11}, \citet{cool13};
(8) \citet{sdssdr7};
(9) \citet{bradshaw13}, \citet{mcclure13};
(10) \citet{delatorre13};
(11) \citet{tasca16};
(12) \citet{lefevre13};
(13) \citet{stalin10}.
\end{table}

\section{Signal-to-noise behaviour}
\label{sec:SN}

\subsection{Signal-to-noise behaviour with magnitude}
\label{sec:ideal}

In Section~\ref{sec:redshift_conditions} we showed that, on average,
we obtain a $Q\ge3$ redshift once a S/N of approximately 2 to 3
in a 1-\AA\ bin is reached.
Here, we inspect how the S/N varies with target magnitude for
a fixed exposure time under uniform observing conditions.  

\begin{figure}
\begin{center}
\includegraphics[width=0.48\textwidth]{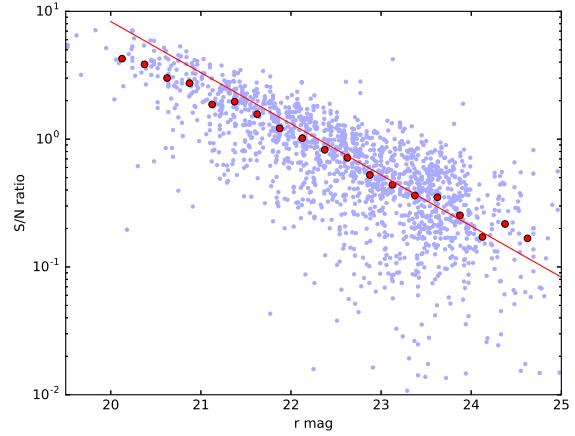}
\caption{Spectrum signal-to-noise (S/N) ratio of SN hosts in 40 minutes of integration plotted against catalog magnitude.  Shown here, as the blue points,  are the results for SN hosts that were observed during the second and third years of OzDES. Plotted in red circles are the median values in bins of 0.25 mag. The red line has a slope of -0.4 (the expectation assuming Poisson statistics) and is fixed to $\mathrm{S/N}=2$ at $m_r\sim21.5$~mag. At bright magnitudes, we expect the red circles to lie below the red line, because of the contribution made by galaxy spectral features to the noise.}
\label{fig:s2n_vs_mag}
\end{center}
\end{figure}

To find frames taken under similar observing conditions, we
use the zeropoints for all single 40-minute OzDES exposures taken
during Y2 and Y3 (the zeropoints for Y1 are less accurate).  These zeropoints
were calculated using the F-star catalogs for the DES-SN fields
\citep[see][for details]{ozdes}.  We select all exposures whose
zeropoints are within 0.1 magnitudes of a chosen value\footnote{The
  chosen value corresponds to data that were taken during the best 10\% of observing conditions.}, 
thereby selecting the frames that
have been observed under a similar observing
conditions. In total, 20 frames were selected. For these exposures we
calculate the S/N of the individual object spectra from 6500 to
8500\AA\ and compare them with the object r-band magnitudes in
Figure~\ref{fig:s2n_vs_mag}. 

We limit the analysis to SN hosts, as they cover a broad range of
magnitudes and a broad range of galaxy types. LRGs, on the other hand,
cover a narrow magnitude range and a narrow range of galaxy types (see
Fig.~\ref{fig:redshift_completeness}).

Figure~\ref{fig:s2n_vs_mag} illustrates that, on average, a S/N of
$\sim 2$ is obtained for SN hosts with $m_r\sim21.5$~mag in a single
40-minute exposure under the best observing conditions.  Hence, from the results presented
in Sec.~\ref{sec:redshift_conditions}, we'd
expect to obtain redshifts for 50\% of sources with $m_r\sim21.5$~mag in a
single exposure.

During a single observing run, most targets are observed with 2 or 3
consecutive 40-minute exposures. Hence, during a single observing run,
we'd expect to obtain redshifts for 50\% of sources with $m_r\sim22$~mag.  In
extrapolating to longer exposures, we have assumed that S/N increases
as $t^{0.5}$. We test this assumption in the next
section.

\subsection{Signal-to-noise accumulation with repeat exposures}
\label{sec:s2n_accumulation}

OzDES uses the AAT to routinely obtain redshifts for
sources as faint $m_{r}\sim24$. Not surprisingly, obtaining
redshifts for sources this faint using a 4-m class telescope requires
long exposures. Some sources have been exposed for
more that a day (1,440 minutes).

These long exposure times give us an opportunity to examine how
the signal-to-noise ratio behaves with exposure time for exposure
times that are rarely reached for fibre fed spectrographs.  With
no sources of systematic error, the signal-to-noise ratio should
increase as $t^{0.5}$. In OzDES, we compute the average spectrum
rather than the sum, so the ideal
case is that the noise in the averaged spectrum decreases with the
square root of the number of exposures. In this section we examine
how close we are to the ideal case.

\begin{figure*}
\begin{center}
\includegraphics[width=0.64\textwidth]{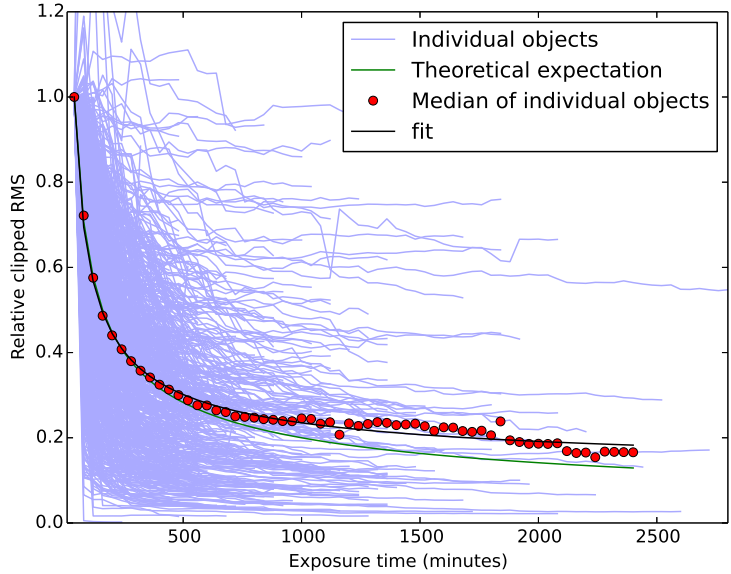}
\caption{The noise in the averaged spectrum for individual objects (red lines) with multiple exposures  as a function of exposure time, normalised to the noise in the first exposure.  Also shown are the median values (black circles) compared to the photon-limited noise expectation (black line) which is proportional to $1/\sqrt{t}$. The blue line is a fit to the black points. Note how the blue line sits slightly above the black line. The horizontal axis has been converted from the number of exposures to the total exposure time by multiplying the former by 40 minutes, which is the exposure time for a single exposure.
}
\label{fig:z_comp_vs_exptime}
\end{center}
\end{figure*}

In Figure~\ref{fig:z_comp_vs_exptime}, we plot how the noise changes
with exposure time.  We computed the noise over the wavelength range 6610 -- 6750\,\AA\, which is a
region that is relatively free of night sky lines.\footnote{We obtain similar
results if we had chosen a region that contains bright night sky
lines.} Each red line in this figure represents a single object. We
exclude F-stars, as the noise may be biased high from real
features in the spectra of these bright objects.

In this figure, there are 800 objects, each containing at least 10
good\footnote{Excludes frames taken when the seeing is worse than
  3\arcsec\ and frames that were taken in cloudy conditions.}
exposures, from the DES C fields, which includes the deep C3 field and the two
shallower C1 an C2 fields \citep[See][for a description of the DES SN fields]{ozdes}. Quantitatively similar results are obtained
with the other 7 DES SN fields. The horizontal axis represents the
number of exposures converted to an exposure time by multiplying the
number of exposures by 40 minutes (the exposure time of a single OzDES
exposure).  The vertical axis has been scaled with respect to the
noise in the first exposure -- hence all blue curves start with a value
of one. As more exposures are added to the averaged spectrum, we
expect the curves to trend downwards.

Ideally, the noise in the spectra that go into making the average is
the same for all exposures. In practice this is not the case because
of variations in seeing and transparency. A natural consequence of these variations
is the large scatter in the behaviour of individual objects in the
figure. The first exposure might have been taken in good (poor)
conditions. If the following exposures were taken in worse (better)
conditions, then the curves will not follow the expected trend with
the number of exposures, which is represented by the black
curve. Thus for this analysis we randomly shuffled the order of the
spectra used in the average, and found that this helps to mitigate
the impact of this variability but does not eliminate it.

The red circles trace out the median behaviour of the individual blue curves
and the solid green line represents the limiting case. The solid black line
is a fit\footnote{We include uncertainties when doing the fit, but for clarity,
we do not show the error bars in Figure~\ref{fig:z_comp_vs_exptime}} to
the red circles using
\begin{equation}
    \frac{n^\alpha + \beta}{1 + \beta}
\label{eq:lower_limit}
\end{equation}
where $n$ is the number of exposures. 

For $\alpha < 0$, the expression approaches $\beta / (1 + \beta)$ as
$n$ becomes large. This represents a lower limit to Eq.~
\ref{eq:lower_limit}. The best fit has $\alpha=-0.6$ and $\beta=0.1$,
which places the lower limit at $\sim 0.09$. The ideal case is
represented by $\beta=0$ and $\alpha=-0.5$. A fit with $\beta$ set to
zero results in $\alpha=-0.48$, which is close to the ideal case, and
is closer to the ideal case than what was found in \cite{sharp10}, who found
$\alpha=-0.32$.

We tested the robustness of this result to changes in the analysis.  If we
used the clipped RMS instead of the normalised median absolute
deviation to compute the noise in each spectrum, or if we used the
zero-points that were determined from F-stars (see Sec.~\ref{sec:inst_setup}) to
scale the spectra instead of the median flux of the spectra themselves,
we found that our results did not change.

As noted above, $\beta=0.1$ corresponds to a lower limit of $\sim
0.09$. In the ideal case, where there is no lower limit, it would take
120 exposures (or 4,800 minutes) for us to reach this level. Note that
none of the objects in Figure~\ref{fig:z_comp_vs_exptime} have been
observed for that many exposures. While one may be tempted to use 120
exposures as the criterion to drop objects, it is likely that OzDES
will soon need to stop observing targets for another reason.  For SN
hosts, we impose a limit of 200 fibres per field per observation in
order to achieve the desired overall balance of OzDES science goals
\citep[see][for futher details]{ozdes}. The other 192 fibres are
assigned to other targets. After 3 years of
observations, we are beginning to reach this limit for the two DES
deep fields (X3 and C3).  Once this limit is reached, we will need to
start removing objects to allocate fibres for newer targets that have a
higher chance of getting a redshift. At the time of writing, the
choice of which objects to remove first has not been resolved.

\section{Forecasts for OzDES and future surveys}
\label{sec:future_surveys}

We use the results from the first half of the OzDES survey to
forecast what will be achieved in the full survey.  In this section we
will focus on the outcomes for the SN host spectroscopy program
within OzDES, since it is the largest source of uniformly selected
targets.  We estimate the final yield of OzDES SN host redshifts
first, and then explore the implications our results have for future
multi-object spectroscopy surveys, particularly in the era of LSST.

\subsection{Forecast for full OzDES SN host redshift yield}

\begin{figure}
\begin{center}
\includegraphics[width=0.48\textwidth]{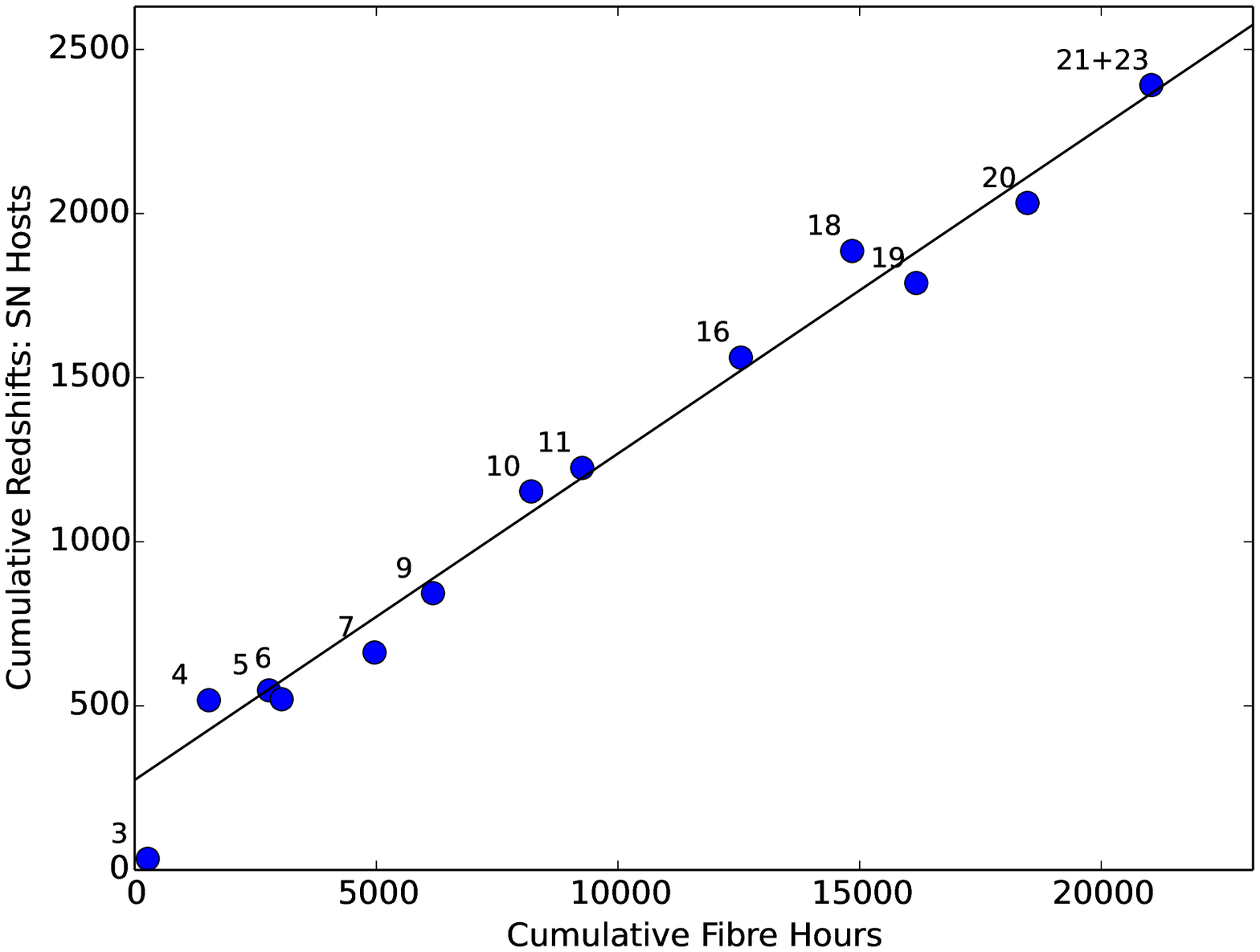}
\includegraphics[width=0.48\textwidth]{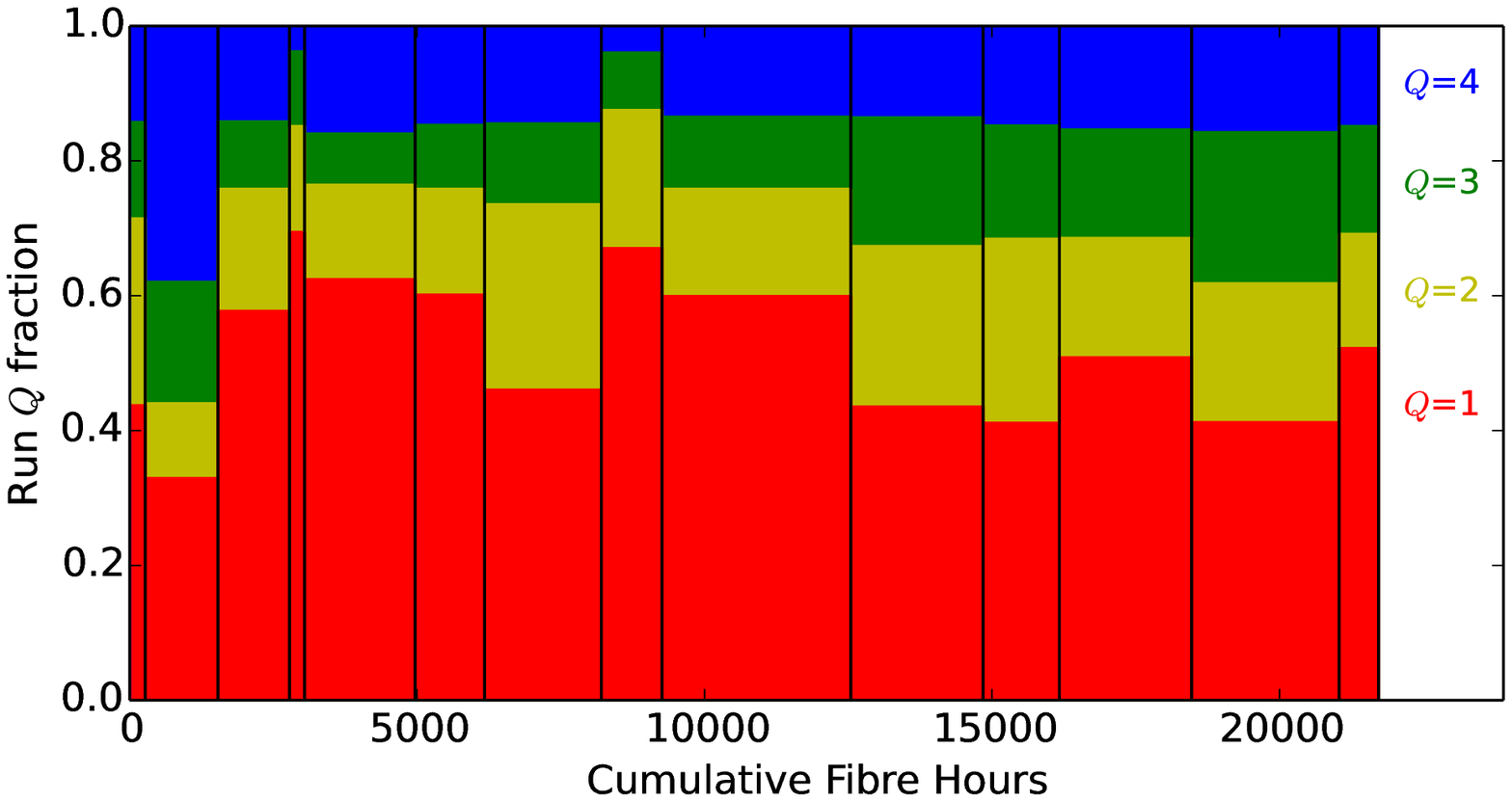}
\caption{Top: Cumulative redshifts versus fibre-hours at the end of
  each OzDES run for SN hosts. Points are labeled according to the
  OzDES observing run number. The first two OzDES runs coincided with the
  science verification phase of DES, and are not included here.
  Bottom: Fraction of observed targets achieving a redshift quality
  $Q$ for the same runs.}
\label{fig:zs_vs_fibhours}
\end{center}
\end{figure}

Here we predict the likely yield of the full five-year OzDES program.
To aid this analysis, we compile the cumulative number of successful
SN host redshifts and the cumulative number of fibre-hours allocated to SN
hosts at the end of each observing run.  We plot these two quantities
in the upper panel of
Figure~\ref{fig:zs_vs_fibhours}.  The lower panel of Figure~\ref{fig:zs_vs_fibhours}
shows the fractions of targets achieving redshift quality $Q$ for each
run, with the width of each run equal to the total additional fibre time
spent on SN hosts. 

These figures show that after the first few observing runs OzDES
quickly achieved a relatively constant rate of accumulation of new
redshifts, shown as the linear fit which has a slope of 0.1 SN host
redshifts per fibre-hour.\footnote{This is equivalent to a 20\%
  success rate per pointing for the default observing strategy of
  3$\times$40 minute exposures.}  This is consistent with a steady state
of new targets added to the queue, a constant fraction of targets
achieving a fully successful $Q=4$ redshift, and the continued
accumulation of S/N for repeat targets.

To calculate the expected SN host redshift yield for the remainder of
the survey, we first must predict the likely future fibre allocation
for SN hosts.  In Y1-Y3, OzDES had 48 allocated observing nights plus
two additional nights, of which 38.3 nights were clear (76\% clear).
OzDES accumulated 92,541 fibre-hours over this period, averaging about
2,400 fibre-hours per night.  Applying these factors (76\% clear
nights\footnote{For Y4, which was in progress as this paper was being
  written, the clear fraction is approximately 75\%.} at 2400 fibre-hours per
night) to the remaining 52 nights allocated to OzDES, we expect around
95,000 fibre-hours remain for the survey.  If SN hosts are allocated
40\% of the remaining fibre-hours\footnote{This is consistent with the
  fraction of fibres that are allocated to SN hosts during the first
  part of Y4.}, this would result in an allocation of 38,000 fibre-hours 
to SN hosts in Y4, Y5 and Y6 of OzDES.

If we apply the steady-state accumulation rate of 0.1 redshifts per
fibre hour to the remaining fibre-hours in the full OzDES survey, then
we would expect an additional 3,800 SN host redshifts. With the \zhost
SN host redshifts already obtained, the final yield would be approximately 6,300
redshifts. However, this is an upper limit. As noted in
Section~\ref{sec:s2n_accumulation}, we are already approaching the
limit of 200 SNe hosts per field for the two deep SN fields C3 and X3,
so in the last three years of the OzDES survey, we may need to remove
SN hosts for which we have been unable to get a redshift, even after
30 hours of integration.

A more detailed simulation, which includes removing targets from the
queue after they have been observed for 30 hours (1800 minutes),
irrespective of their status, and includes the probability of
obtaining a redshift as a function of exposure time (see
Fig.~\ref{fig:completeness_mag}), results in a yield of 5,700 SN host
redshifts. This is our current best estimate, but it depends on the
number of clear nights we obtain during the second half of the survey.

\subsection{Insights for future MOS surveys}
\label{sec:other_mos_future}
Finally, we turn to the prospects for future MOS surveys in light of
the insights gained from the first three years of OzDES. Here, we
will focus on future MOS facilities that are capable
of devoting some portion of their fibres toward the
observation of supernova host-galaxies, particularly for
SNe discovered by the Large Synoptic Survey Telescope (LSST).  We will
analyse the potential role these facilities could play in a parallel
role for LSST as that served by OzDES for DES: classifying live
transients and acquiring host-galaxy redshifts for photometric SN
classification.

We will primarily focus on three future facilities: the 4-metre
Multi-Object Spectroscopic Telescope \citep[4MOST;][]{4most}, the
Subaru Prime Focus Spectrograph \citep[PFS;][]{sugai12, takada14},
and the Maunakea Spectroscopic Explorer \citep[MSE;][]{mse1, mse2}.
Numerous other facilities, both in existence and planned for future
construction, have MOS capability and will conduct MOS surveys in the
LSST era \citep[see][for a summary]{mse2}.  These include DESI
\citep{desi}, which will focus on obtaining galaxy redshifts for Baryon
Acoustic Oscillation (BAO) science, WEAVE \citep{weave}, which will
focus on followup of Gaia stars and LOFAR radio galaxies, and MANIFEST
\citep{Lawrence2016} on the Giant Magellan Telescope. 

\begin{table*}
\begin{minipage}{0.8\textwidth}
\begin{center}
\caption{Properties of select current and forthcoming MOS facilities.}
\label{tab:mos_facilities}
\begin{tabular*}{\textwidth}{@{\extracolsep{\fill}}lcccc}
\hline
 & AAT & 4MOST* & PFS\# & MSE$\dag$ \\
\hline
{\em Facility} & & & & \\
\,\, - Aperture      & 3.9m        &  4.1m           & 8.2m         & 11.25m \\
\,\, - Field of View & 3.4 deg$^2$ &  4.1 deg$^2$      & 1.25 deg$^2$ & 1.5 deg$^2$ \\
 & & & \\
{\em (Low-Res.) Spectrograph} & & & & \\
\,\, - Resolution              & 1800         & 4000-7800    & 2300-5000          & 2500-3000  \\
\,\, - Wavelength Range (\AA)  & 3800-8900    &  3700-9500   & 3800-12600         & 3600-18000   \\
\,\, - Fibre Diameter          & 2.0 \arcsec\ & 1.45\arcsec\ & 1.1\arcsec\  & 1.2\arcsec\ \\
\,\, - Fibre Multiplex         & 400          &  1624        & 2394               & 3200 \\
\hline
\multicolumn{5}{l}{*for the Low Resolution Spectrograph, from \footnotesize{https://www.4most.eu/cms/facility/}}\\
\multicolumn{5}{l}{\# \footnotesize{\cite{takada14}}}\\
\multicolumn{5}{l}{$\dag$ \footnotesize{http://mse.cfht.hawaii.edu/docs/mse-science-docs/DSC/MSE\_Precis\_v8.pdf}}\\
\end{tabular*}
\end{center}
\end{minipage}
\end{table*}

We summarise the key properties of these three facilities in
Table~\ref{tab:mos_facilities}.  4MOST has similar characteristics to
AAOmega in terms of wavelength coverage and telescope aperture. The
ratio of the fibre size to average site seeing is also similar.  For
point sources, the gain from the combination of better seeing and
smaller fibre size is as large as the gain one obtains from a larger
telescope aperture. For constant fibre size and constant seeing, the
signal-to-noise ratio goes as the aperture of the telescope under
background-limited conditions. For constant aperture and a constant
ratio between fibre size and site seeing, the signal-to-noise ratio
for point sources goes as the inverse of the diameter of the
fibre. For extended sources, the improvement will be slower. Computing
the signal-to-noise ratio for extended sources requires an accurate
model of the size distribution of SN host-galaxies. This is beyond the
scope of the current paper, and we leave this to future work.

We expect 4MOST on VISTA will be a factor of two faster than AAOmega
on the AAT at obtaining the redshift of the same object, with
potential added benefits that come from higher spectral resolution,
which reduces the impact from night sky lines at red wavelengths, and
broader wavelength coverage.  Similarly, we expect that PFS and MSE
will be 15 and 23 times faster than AAOmega, respectively.  Ignored in
these comparisons are differences in instrument throughput, which one
might expect to be better for future instruments.

Taking into account the patrol area of these instruments, the area
these instruments will be able to cover in the same amount of
observing time will be 2.4, 5.2 and 10 times that of the AAT, for 4MOST,
PFS, and MSE, respectively.

In computing the likely number of host-galaxy redshifts from these
surveys, one needs to make assumptions about the observing strategies
of the MOS surveys, and the observing strategy of LSST. Under the
assumption that the observing strategies of the MOS surveys are the
same as OzDES and that the distribution of SN host-galaxy magnitudes
from LSST is the same as DES\footnote{The redshift distribution of SNe
  from DES and LSST are broadly comparable
  \citep[cf. Fig.~\ref{fig:redshift_completenessdistribution} with
    Fig. 11.1 in][]{LSST2009}.}, then for a SN host-galaxy survey with
4MOST using 60,000 fibre-hours (the same number of hours OzDES will
spend on hosts by the end of its survey), we would expect a yield of
the order 14,000 SN host-galaxy redshifts, which is almost a factor of
two larger than the expected yield from OzDES. For PSF and MSE, the
numbers of redshifts are 30,000 and 57,000, respectively. Here the
increase comes from being able to cover a larger area of the sky in
the same observing time.

Not all supernovae with a host redshift will be placed on the Hubble
diagram and used in fitting the cosmology. At the end of 3yrs,
approximately 40\% of DES supernova with a host redshift remain after
light curve sampling and light curve quality cuts are
applied. Assuming that a similar fraction are kept in future surveys,
one can expect a yield of 5,600, 12,000 and 23,000 SNe Ia for 4MOST, PSF, and
MSE, respectively.

However, the number of SNe Ia will be further constrained by the
observing strategy of LSST. Similarly to DES,
LSST will observe a number of deep fields \citep{LSST2009}. In one
realisation of the LSST survey, approximately 1,500
SNe Ia with light curves of sufficient quality in at least three
filters will be obtained per year from 10 LSST deep fields. Over the 10
years that LSST will run, this comes to 15,000 SNe Ia. This then
limits the number of SNe Ia that can be placed on the Hubble diagram
from an OzDES-sized survey using MSE.

As further details on the observing strategies of LSST and 
future MOS facilities emerge, the results from OzDES will make it possible to
provide more precise estimates of the yields of these
upcoming facilities.

\section{Conclusions}
\label{sec:conclusions}
In this work, we presented the results of the first three years of
OzDES, a spectroscopic survey obtaining redshifts in the 10 DES SN
fields.  Over 52 observing nights on the AAT, OzDES has obtained
\ztotalAll redshifts in the DES supernova fields.  These include a wide
variety of targets, such as supernova host-galaxies, AGN, LRGs, radio
galaxies, and numerous photometric redshift training samples. At the
same time, OzDES has spectroscopically confirmed almost 100 SNe.  This
work also marks the first OzDES data release (OzDES-DR1), which sees
the release of \ztotal redshifts (we note that the SN host redshifts
will be released in a future DES analysis).

We examined the requirements for obtaining a redshift in the OzDES
data.  For nearly all target types (excepting only emission line
galaxies), we obtain a redshift for 50\% of sources that have a
signal-to-noise of 2 to 3 in bins of 1\AA\ in the red part of the
spectrum (observer frame 6,500-8,500 \AA).  

We also examined the behaviour of the signal-to-noise ratio with
exposure time, finding that the change in the signal-to-noise ratio
with exposure time closely matches the Poisson limit for exposures as
long as 10 hours, which is a measure of how well we control systematic
errors in the data processing. However, for longer exposures, the
signal-to-noise starts to depart from the Poisson limit.

We use these results to estimate the redshift yields for the
remainder of the OzDES survey or other similar future surveys. With
the proviso that observing conditions will be similar to what we've
experienced so far, we predict a final yield of 5,700 SN host
redshifts.

\vskip11pt
{\em Acknowledgements:}
This research was conducted by the Australian Research Council Centre of Excellence for All-sky Astrophysics (CAASTRO), through project number CE110001020. BPS acknowledges support from the Australian Research Council Laureate Fellowship Grant FL0992131. R.J.F. is supported in part by fellowships from the Alfred P. Sloan Foundation and the David and Lucile Packard Foundation.
Based in part on data collected at the Australian Astronomical Observatory, through program A/2013B/012.
This research has made use of the NASA/IPAC Extragalactic Database (NED) which is operated by the Jet Propulsion Laboratory, California Institute of Technology, under contract with the National Aeronautics and Space Administration.
This research has made use of NASA's Astrophysics Data System (ADS).
We thank the anonymous referee for very helpful feedback which improved the quality of this paper.

Funding for the DES Projects has been provided by the U.S. Department of Energy, the U.S. National Science Foundation, the Ministry of Science and Education of Spain, 
the Science and Technology Facilities Council of the United Kingdom, the Higher Education Funding Council for England, the National Center for Supercomputing 
Applications at the University of Illinois at Urbana-Champaign, the Kavli Institute of Cosmological Physics at the University of Chicago, 
the Center for Cosmology and Astro-Particle Physics at the Ohio State University,
the Mitchell Institute for Fundamental Physics and Astronomy at Texas A\&M University, Financiadora de Estudos e Projetos, 
Funda{\c c}{\~a}o Carlos Chagas Filho de Amparo {\`a} Pesquisa do Estado do Rio de Janeiro, Conselho Nacional de Desenvolvimento Cient{\'i}fico e Tecnol{\'o}gico and 
the Minist{\'e}rio da Ci{\^e}ncia, Tecnologia e Inova{\c c}{\~a}o, the Deutsche Forschungsgemeinschaft and the Collaborating Institutions in the Dark Energy Survey. 

The Collaborating Institutions are Argonne National Laboratory, the University of California at Santa Cruz, the University of Cambridge, Centro de Investigaciones Energ{\'e}ticas, 
Medioambientales y Tecnol{\'o}gicas-Madrid, the University of Chicago, University College London, the DES-Brazil Consortium, the University of Edinburgh, 
the Eidgen{\"o}ssische Technische Hochschule (ETH) Z{\"u}rich, 
Fermi National Accelerator Laboratory, the University of Illinois at Urbana-Champaign, the Institut de Ci{\`e}ncies de l'Espai (IEEC/CSIC), 
the Institut de F{\'i}sica d'Altes Energies, Lawrence Berkeley National Laboratory, the Ludwig-Maximilians Universit{\"a}t M{\"u}nchen and the associated Excellence Cluster Universe, 
the University of Michigan, the National Optical Astronomy Observatory, the University of Nottingham, The Ohio State University, the University of Pennsylvania, the University of Portsmouth, 
SLAC National Accelerator Laboratory, Stanford University, the University of Sussex, Texas A\&M University, and the OzDES Membership Consortium.

The DES data management system is supported by the National Science Foundation under Grant Number AST-1138766.
The DES participants from Spanish institutions are partially supported by MINECO under grants AYA2015-71825, ESP2015-88861, FPA2015-68048, SEV-2012-0234, SEV-2012-0249, and MDM-2015-0509, some of which include ERDF funds from the European Union. IFAE is partially funded by the CERCA program of the Generalitat de Catalunya.

\bibliographystyle{apj}
\bibliography{ozdes_y3}

\appendix
\section{Observing Logs}
\label{app:observing_logs}
The observing runs logs for OzDES Y2 and Y3 are presented below.

\begin{table*}
\centering
\begin{minipage}{\textwidth}
\caption{OzDES second year (Y2) observing log for DES SN fields \citep[c.f. Table 2 of][]{ozdes}.}
\label{tab:observing_log_y2}
\begin{tabular*}{\textwidth}{@{\extracolsep{\fill}}ccccccccccccc}
\hline
UT Date & Observing Run\textsuperscript{a} & \multicolumn{10}{c}{Total exposure time for DES field (minutes)} & Notes for entire run \\
& & E1 & E2 & S1 & S2 & C1 & C2 & C3(deep) & X1 & X2 & X3(deep)\\
\hline
\\ 
2014-09-17 &  007  & 120 & 120 & --  & --  & --  & --  & 110 & --  & --  & 120 & Half night lost to weather.\\
2014-09-18 &       & 120 & --  & --  & 120 & --  & --  & --  & --  & 120 & --  & \\
2014-09-19 &       & --  & 120 & 120 & 40  & --  & --  & --  & 120 & --  & --  & \\
2014-09-20 &       & 160 & --  & --  & --  & --  & 120 & --  & --  & --  & 120 & \\
2014-09-21 &       & 60* & 50* & 80  & --  & 160 & --  & --  & --  & --  & --  & \\
\\ 
2014-10-27 &  009  & --  & --  & --  & --  & 75* & 75* & 50* & --  & --  & 75* & Half night lost to instrument error.\\
2014-10-28 &       & 90  & 90  & --  & --  & --  & --  & --  & --  & 90  & 90  & \\
2014-10-29 &       & --  & --  & --  & --  & --  & 70  & 90  & --  & --  & --  & \\
2014-10-30 &       & --  & --  & --  & --  & 90  & --  & --  & 90  & --  & --  & \\
\\ 
2014-11-18 &  010  & --  & 120 & --  & --  & --  & --  & 120 & --  & --  & 120 & Half night lost to weather.\\
2014-11-19 &       & 120 & --  & --  & --  & 80  & 90  & --  & 120 & --  & --  & \\
2014-11-20 &       & --  & --  & 40* & 80* & --  & --  & --  & --  & 80* & --  & \\
2014-11-21 &       & --  & --  & 120 & --  & --  & --  & 80  & --  & 120 & --  & \\
2014-11-27 &       & --  & --  & --  & --  & --  & --  & --  & --  & --  & 80  & \\
\\ 
2014-12-21 &  011  & --  & --  & --  & --  & 40* & --  &120* & --  & --  & 80  & Four nights lost to weather\\
2014-12-22 &       & --  & --  & --  & --  & --  & --  & --  & --  & --  & --  & \\
2014-12-23 &       & --  & --  & --  & --  & --  & --  & --  & --  & --  & --  & \\
2014-12-24 &       & --  & --  & --  & --  & 90  & 80  & --  & --  & --  & 120 & \\
2014-12-25 &       & --  & --  & --  & --  & --  & --  & --  & --  & --  & --  & \\
2014-12-26 &       & 80* & --  & --  & --  & --  & --  & --  & --  & --  & --  & \\
2014-12-29 &       & --  & 80  & --  & --  & 80  & --  & --  & --  & --  & --  & \\
2014-12-30 &       & 80  & --  & --  & --  & --  & --  & --  & --  & --  & --  & \\
\\
2014 (All Y2) & Total (min)
                   & 690 & 530 & 320 & 160 & 500 & 360 & 400 & 330 & 330 & 730 & \\
\hline
\\
\multicolumn{13}{l}{\textsuperscript{a}\footnotesize{Run numbering includes runs from 2dFLenS project, so is not necessarily contiguous.}}\\
\multicolumn{13}{l}{* \footnotesize{Bright object backup programme for poor conditions, not counted toward final total.}}\\
\end{tabular*}
\end{minipage}
\end{table*}

\begin{table*}
\centering
\begin{minipage}{\textwidth}
\caption{OzDES third year (Y3) observing log for DES SN fields \citep[c.f. Table 2 of][]{ozdes}.}
\label{tab:observing_log_y3}
\begin{tabular*}{\textwidth}{@{\extracolsep{\fill}}ccccccccccccc}
\hline
UT Date & Observing Run\textsuperscript{a} & \multicolumn{10}{c}{Total exposure time for DES field (minutes)} & Notes for entire run \\
& & E1 & E2 & S1 & S2 & C1 & C2 & C3(deep) & X1 & X2 & X3(deep)\\
\hline
\\ 
2015-08-07 &  016  & 40  & --  & --  & --  & --  & --  & 90  & --  & --  & --  & 3.5/4.0 clear nights\\
2015-08-08 &       & 80  & --  & --  & --  & --  & --  & 60  & --  & --  & --  & \\
2015-08-09 &       & --  & 120 & --  & --  & --  & --  & 80  & --  & --  & 120 & \\
2015-08-10 &       & 120 & --  & --  & --  & 120 & --  & --  & --  & 80  & --  & \\
2015-08-11 &       & --  & 120 & 130 & --  & --  & 80  & --  & --  & --  & --  & \\
2015-08-18 &       & 120 & --  & --  & 70  & --  & --  & --  & 80  & --  & --  & \\
\\ 
2015-09-16 &  018  & 80  & 80  & --  & --  & --  & 70  & 80* & --  & --  & 50  & 4.0/4.5 clear nights\\
2015-09-17 &       & 80  & 40  & --  & --  & --  & 40  & 50  & --  & --  & 80  & \\
2015-09-18 &       & --  & --  & --  & --  & 110 & --  & --  & 80  & 120 & --  & \\
2015-09-19 &       & 120 & --  & 80  & 90  & --  & --  & --  & --  & --  & 80  & \\
2015-09-20 &       & --  & 70* & 80  & --  & --  & --  & 80  & 80  & --  & --  & \\
2015-09-21 &       & 10* & --  & --  & 90  & --  & --  & --  & --  & 70* & --  & \\
\\ 
2015-10-10 &  019  & 80  & 80  & --  & --  & --  & --  & 80  & --  & --  & 120 & 2.5/4.0 clear nights\\
2015-10-11 &       & 80  & 80  & --  & --  & --  & --  & --  & 40  & --  & --  & \\
2015-10-12 &       & --  & --  & --  & --  & --  & --  & --  & --  & --  & --  & \\
2015-10-13 &       & 80  & 80  & 80  & --  & --  & --  & 80  & 120 & --  & --  & \\
\\ 
2015-11-12 &  020  & --  & --  & --  & --  & 50* & --  & 100 & --  & --  & --  & 3.0/4.0 clear nights\\
2015-11-13 &       & --  & --  & --  & --  & --  & 140 & 80  & --  & --  & --  & \\
2015-11-14 &       & 80  & --  & --  & --  & --  & --  & --  & --  & --  & 120 & \\
2015-11-15 &       & --  & 40  & --  & --  & 100 & --  & --  & 120 & 120 & --  & \\
\\ 
2015-12-03 &  021  & --  & --  & 120 & 120 & --  & --  & --  & --  & --  & --  & 3.1/4.0 clear nights\\
2015-12-04 &       & --  & --  & --  & --  & --  & --  & --  & --  & --  & 60  & \\
2015-12-12 &       & --  & 80  & --  & --  & --  & --  & 120 & --  & --  & 120 & \\
2015-12-13 &       & 80  & --  & --  & --  & 120 & --  & --  & 80  & --  & --  & \\
2015-12-14 &       & --  & 80  & --  & --  & --  & 80  & --  & --  & 120 & --  & \\
2015-12-15 &       & --  & --  & --  & --  & --  & --  & --  & --  & --  & --  & \\
\\ 
2016-02-06 &  023  & --  & --  & --  & --  & --  & --  & 110 & --  & --  & --  & 0.7/0.7 clear nights\\
2016-02-07 &       & --  & --  & --  & --  & --  & 120 & --  & --  & --  & --  & \\
\\
2015 (All Y3) & Total (min)
                   & 920 & 800 & 490 & 370 & 450 & 530 & 930 & 600 & 440 & 750 & \\
\hline
\\
\multicolumn{13}{l}{\textsuperscript{a}\footnotesize{Run numbering includes runs from 2dFLenS project, so is not necessarily contiguous.}}\\
\multicolumn{13}{l}{* \footnotesize{Bright object backup programme for poor conditions, not counted toward final total.}}\\
\end{tabular*}
\end{minipage}
\end{table*}

\end{document}